\documentclass[aps,pre,preprint,groupedaddress]{revtex4-1}
\usepackage[latin1]{inputenc}
\usepackage{amsmath}
\usepackage{amsfonts}
\usepackage{amssymb}
\usepackage{graphicx}
\usepackage[colorlinks=true, citecolor=red, linkcolor=blue,urlcolor=blue ]{hyperref}
\usepackage{bm}
\usepackage{subcaption}

\renewcommand{\u}{\bm{u}}
\renewcommand{\r}{\bm{r}}
\newcommand{\e}{\bm{e}}
\newcommand{\n}{\bm{n}}
\newcommand{\bPi}{\bm{\Pi}}
\newcommand{\W}{\bm{W}}
\newcommand{\Da}{\rm{Da}}

\begin{document}

	\title{Stochastic dynamics of dissolving active particles}
			
	\author{Alexander Chamolly}
	\author{Eric Lauga}
	\email{e.lauga@damtp.cam.ac.uk}
	\affiliation{Department of Applied Mathematics and Theoretical Physics, University of Cambridge, Wilberforce Road, Cambridge CB3 0WA, United Kingdom}
	
	\date{\today}

	\begin{abstract}
		The design of artificial microswimmers has generated significant research interest in recent years, for promise in applications such as nanomotors and targeted drug-delivery. However, many current designs suffer from a common problem, namely  the swimmers remain in the fluid indefinitely, posing risks of clogging and damage. Inspired by recently proposed experimental designs, we investigate mathematically the dynamics of  degradable active particles. We develop and compare  two distinct chemical models for the decay of a swimmer, taking into account the material composition and nature of the chemical or enzymatic reaction at its surface. These include a model for dissolution without a reaction, as well as models for a reacting swimmer studied in the limit of  large and  small Damk\"ohler number. 
		A new dimensionless parameter emerges that allows the classification of colloids into ballistic and diffusive type. Using this parameter, we perform an asymptotic analysis to derive expressions for colloid lifetimes and their total mean-squared displacement from release and validate these by numerical Monte Carlo simulations of the associated Langevin dynamics. Supported by  general scaling relationships, our theoretical results provide new insight into the experimental applicability of a wide range of designs for degradable active colloids.

	\end{abstract}	

	\maketitle
	
	\newpage
	
	\section{Introduction}
	In recent years, scientists from a wide  variety of different fields have given considerable  attention to the subject of synthetic microswimmers. This focus in research is no coincidence, as such colloids show great promise in biomedical and engineering applications \cite{wang2012nano,wang2013small,nelson2010microrobots}. The design of autonomous swimmers in particular has received significant theoretical and experimental attention \cite{elgeti2015physics,moran2017phoretic}. In an effort to exploit the peculiarities of the associated low-Reynolds number hydrodynamics \cite{purcell1977life}, many different propulsion mechanisms have been invented. These include self-phoretic propulsion, such as chemophoresis \cite{michelin2014phoretic,golestanian2007designing,brady2011particle,walther2013janus} and electrophoresis \cite{ebbens2014electrokinetic,paxton2006catalytically,moran2011electrokinetic}, as well as ultrasound propulsion \cite{gallino2018physics,mou2015single,wang2012autonomous}, bubble propulsion \cite{gibbs2009autonomously,wang2014selecting} and magnetic propulsion \cite{zhang2009characterizing,ghosh2009controlled}.
	
	Despite this remarkable progress, common experimental designs still need to be improved in order to be suitable for sensitive applications, such as non-invasive medicine. Next to potential toxicity of swimmer components or their fuel \cite{gao2015artificial}, the question of waste disposal remains largely open. This can be a serious problem, since artificial micron sized particles in the blood stream have the potential to cause clogging \cite{bacher2017clustering,sauret2018growth,fogelson2015fluid} and may thus pose a significant health risk \cite{nesbitt2009shear,fogelson2015fluid}. It is therefore essential to develop designs for microswimmers that degrade after fulfilling their purpose.
	
	 Very recently, novel experimental designs have begun to address these issues. Examples of such colloids include non-toxic magnesium-based bubble propelled swimmers \cite{chen2018magnesium} suitable for aqueous environments, as well as other kinds of inorganic compositions driven by reactions in either acidic or alkaline environments \cite{chen2016transient}. More designs have been proposed using organic compounds that may be 3D-printed   \cite{wang20183d} or that self-assemble into nanomotors \cite{tu2017biodegradable}.
	
	These experimental advances raise new theoretical questions. While the  dynamics of classical non-dissolving colloids have been studied extensively, the time-evolution of colloid size modifies its stochastic behaviour, and new quantities characterising its physics emerge. The purpose of this paper is therefore to provide theoretical answers to two fundamental questions. First, we examine which material and environmental parameters determine the lifetime of a dissolving spherical microswimmer. Second, we study the influence of dissolution on the stochastic behaviour of both passive and self-propelled colloids. Here, a new dimensionless quantity arises which splits microswimmers into two categories: those that are subject to sufficient amounts of thermal noise during their life time to evolve diffusively, and those that exhibit near-ballistic trajectories that may be exploited for delivery applications. We show that both scenarios may enter for realistic values of the material and environmental parameters. Knowledge of these and their scaling relations is thus essential for the application-specific engineering of degradable microswimmer designs.
	
	The structure of this paper is as follows. We begin by presenting two theoretical models for the dissolution process in \S \ref{sec:Models}, one suitable for designs in which the dissolution process is not driven by a reaction with a fuel in the solvent (such as dissolution by hydrogen bonding), and one for swimmers whose matrix is decomposed by means of a reaction (chemical or enzymatic). For further analysis the latter case is considered in the two limits of   slow and   fast reaction, the former corresponding to a fixed material flux boundary condition. In all these models we find expressions for the time dependence of the swimmer size, as well as their total lifetime in terms of the essential physical parameters. We present the necessary modification to classical Brownian motion in \S \ref{sec:Passive}, and derive expressions for the passive mean squared displacement of not self-propelling colloids. Based on this, we next derive corresponding expressions for active motion in \S \ref{sec:Active} and validate our results numerically. Finally we discuss the implications of our research on future studies in \S \ref{sec:Discussion}.

\section{Dissolution models}\label{sec:Models}
Inspired by recent experimental realisations, we propose two models for the dissolution of a spherical colloid based on different possibilities for the boundary conditions at its surface. Specifically, we distinguish between the case in which dissolution occurs through binding colloid material to fluid molecules (for example, the case of ionic dissolution in water), which we call non-reacting, and the case of dissolution through a chemical or enzymatic reaction that consumes a fuel. In the latter scenario we distinguish further between the  limits of   slow and   fast reaction, and discuss their physical implications.

As a preamble, we note that, unlike geophysical melting processes~\cite{woods1992melting}, enthalpy plays no role in the dissolution processes considered in our paper. This means the Stefan boundary condition does not apply and the dynamics we derived is different from e.g.~the dissolution of ice crystals in water. While the general dynamics of diffusive dissolution have been considered in the geophysical literature~\cite{zhang1989diffusive}, there has to the best of our knowledge been no study that derived the asymptotic solutions we compute below. This is likely due to the dominance of convection driven processes on relevant geophysical scales that require different modelling~\cite{kerr1995convective}.

\subsection{Non-reacting swimmer}

\begin{figure}[t!]
	\centering
	\includegraphics[width=0.5\columnwidth]{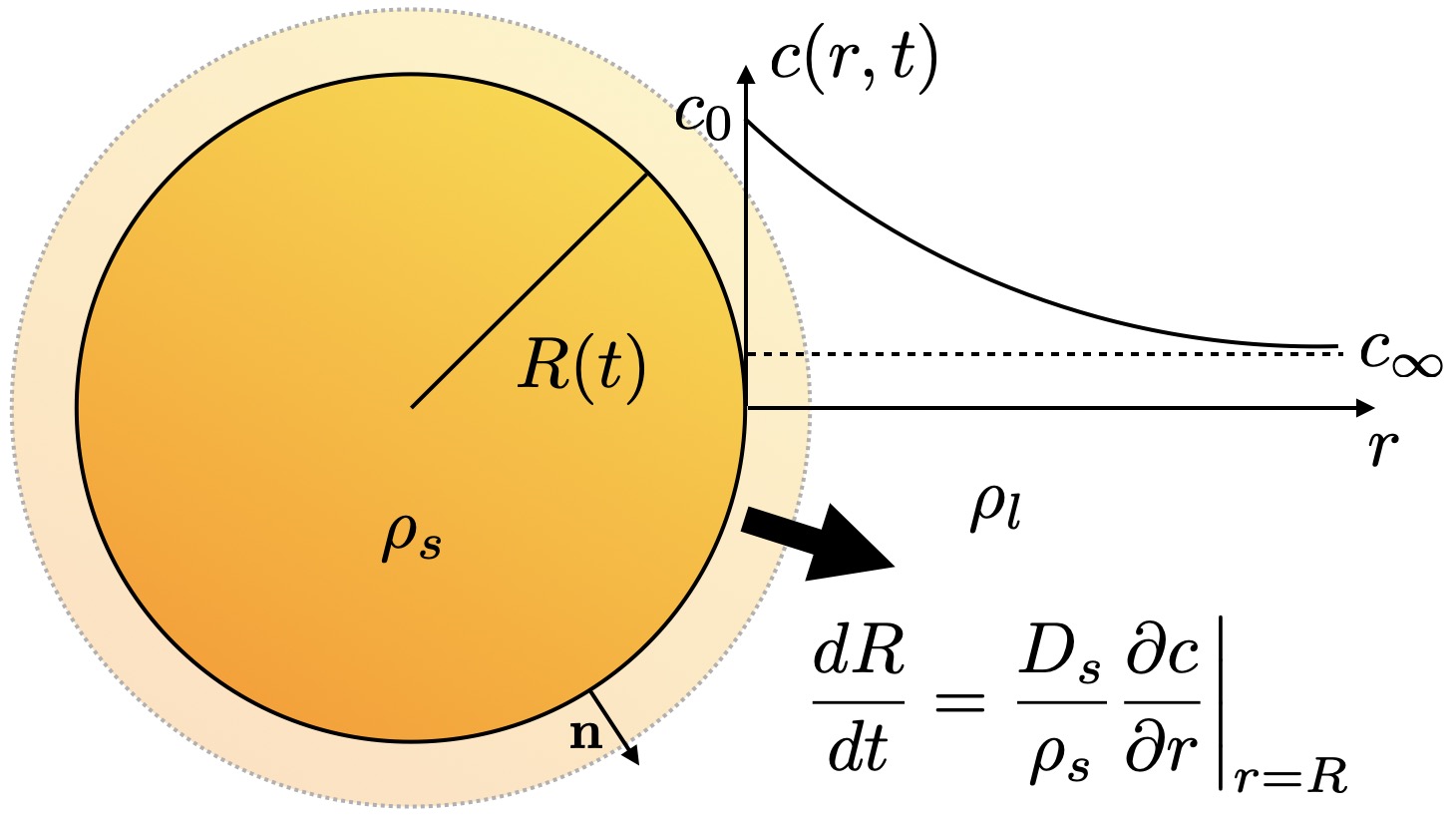}
	\caption{Schematic presentation of non-reacting dissolution dynamics. The matrix of the swimmer consists of a substance that dissolves by bonding to the fluid (thus acting as a solvent). Near the boundary, solute is present at a saturation concentration $c_0$ and subject to advective-diffusive transport in the bulk. Dissolution emerges through maintaining a normal concentration gradient at the swimmer surface.}\label{fig:glreac}
\end{figure}

In our first model, we assume that the colloidal particle is composed of a material that dissolves in the surrounding fluid through bonding of solute colloid material to fluid molecules, as illustrated schematically in Fig.~\ref{fig:glreac}. We consider this an appropriate model for non-reacting dissolution processes, such as dissolution of many organic compounds as well as ionic salts in water. In order to keep the mathematics simple we make the simplifying assumption that only one species of solute is dissolved into the bulk. This allows us to define the (mass) concentration, $c(\r,t)$, of solute defined as the mass of solute dissolved in a unit volume of solvent, with $c=c_\infty\geq0$ far away from the colloid. Note that this differs from the definition of molar concentration common in chemistry by a factor equal to the molar mass of the solute. We make this choice in order to avoid clutter that would arise from the application of mass conservation below.

In this model and the following we assume the absence of any background flow that would disturb the distribution of solute or reactant in the bulk fluid. This assumption is of course violated for self-propelled particles moving relative to a background fluid. However, we can use a scaling argument to show that this does not affect our leading-order results. Since typical propulsion velocities $U$ are expected to be on the order of a few microns per second, initial colloid radii $R_0$ on the scale of microns \cite{elgeti2015physics} and for many ions in water at room temperature the solute diffusivity is approximately $D_s\sim10^{-9}$~m$^2$/s \cite{haynes2014crc}, the P\'eclet number quantifying the relative important of advection to diffusion for the solute is $\text{Pe}_\text{sol}=R_0U/D_s\sim10^{-4}-10^{-3}$. This indicates that advection of solute can be safely neglected. This remains true even when the P\'eclet number associated with motion of the colloid, $\text{Pe}_\text{col}=R_0U/D$ is large, since the particle is several orders of magnitude larger than a solvent molecule and therefore has a much smaller diffusivity. The same result applies to phoretic slip flows, which are typically of the same strength as the propulsion velocity. In the context of dissolution dynamics, the flows arising from propulsion can therefore be neglected in the transport processes of solute and reactant.

We further assume that the swimmer has a homogeneous mass density, $\rho_s$, and the fluid solvent a constant density, $\rho_l$. In general the density of the solvent depends weakly on the amount of solute dissolved \cite{haynes2014crc}. However, we will soon develop an asymptotic analysis based on the assumption that the solubility is weak and therefore can neglect this effect. Finally, we assume also that the swimmer remains spherical at all times, and that that the dissolution dynamics is independent of any self-propulsion mechanism or background flow. Both these assumptions will be justified \emph{a posteriori} in section \S\ref{sec:physint}. A brief discussion of the case of a partially dissolving swimmer is included our discussion \S\ref{sec:Discussion}.

\subsubsection{Mathematical model}

We consider a spherically symmetric colloid or radius $R(t)$ with initial condition $R(0)=R_0>0$. Near the boundary, there is  chemical equilibrium between solute attached to the swimmer surface and present in the fluid. In this case the dissolution process is driven by removal (through   diffusion) of solute from a boundary layer into the bulk and subsequent replenishment from the swimmer surface (Fig.~\ref{fig:glreac}). We model this effect by imposing the boundary condition
\begin{equation}
c(R(t),t)=c_0>c_\infty,\quad t\geq 0,
\end{equation} 
where $c_0$ is the saturation concentration of solute in the solvent. This condition assumes that the boundary layer is negligibly thin and that the surface reaches chemical equilibrium instantaneously, which may be justified by noting that time scales of interest will be much larger than the molecular collision time, $\tau_{MC}\approx 10^{-13}\text{ s}$ \cite{haynes2014crc}. The other condition we impose is the requirement that the solute is initially distributed homogeneously in the bulk, i.e.
\begin{equation}
c(r,0)=c_\infty,\quad r>R_0.
\end{equation}
Conservation of solute at the boundary gives
\begin{align}
4\pi R^2 \rho_s \frac{dR}{dt}&=-\text{(solute flux into the fluid)}\nonumber\\
&=-\left(-D_s 4\pi R^2 \frac{\partial c}{\partial r}\bigg\rvert_{r=R} \right),
\end{align}
and therefore
\begin{equation}\label{eq:3solbc}
 \frac{dR}{dt}= \frac{D_s}{\rho_s}\frac{\partial c}{\partial r}\bigg\rvert_{r=R},
\end{equation}
where $D_s$ is the diffusivity of solute in the solvent. 

Furthermore, in the case of unequal densities we also get a non-zero fluid flux at the boundary since by mass conservation there is equality
\begin{equation}
-\dot{R}\rho_s = (-\dot{R}+{\u}\cdot{\hat{\r}})\rho_l
\end{equation}
and thus
\begin{equation}\label{eq:uBC}
{\u}\cdot{\hat{\r}} = \dot{R}\frac{\rho_l-\rho_s}{\rho_l},
\end{equation}
where $\hat{\r}$ denotes a unit vector in the outward radial direction.

For a self-propelled microscopic colloid in water the Reynolds number, defined as the ratio of colloid radius times velocity divided by kinematic viscosity, is typically on the order of $10^{-9}\ll 1$. Therefore the fluid dynamics obey  the  incompressible Stokes equations,
\begin{equation}
	\mu\nabla^2\u =\nabla p,\quad \nabla\cdot \u=0,
\end{equation}
where $\mu$ is dynamic viscosity and $p$ is the pressure field. Solving these with the boundary condition given in Eq.~\eqref{eq:uBC} at $r=R$ leads to the flow of a point source
\begin{equation}\label{eq:flow}
\u=\dot{R}\frac{\rho_l-\rho_s}{\rho_l}\frac{R^2}{r^2}\hat{{\r}}.
\end{equation}
The transport equation for $c(r,t)$ is the standard  advection-diffusion equation
\begin{eqnarray}
\frac{\partial c}{\partial t}+\nabla\cdot (c\u)=D_s\nabla^2c.
\end{eqnarray}
Using the result of Eq.~\eqref{eq:flow} together with incompressibility and assuming radial symmetry of the solute concentration, this becomes 
\begin{align}\label{eq:transport}
  \frac{\partial c}{\partial t} + \frac{\rho_l-\rho_s}{\rho_l}\frac{D_s}{\rho_s}\frac{R^2}{r^2}\frac{\partial c}{\partial r}\bigg\rvert_{r=R}\frac{\partial c}{\partial r}= D_s \left(\frac{\partial^2 c}{\partial r^2}+\frac{2}{r}\frac{\partial c}{\partial r}\right),
\end{align}
Next we non-dimensionalise this transport equation using the scalings
\begin{equation}
c^*=\frac{c-c_\infty}{c_0-c_\infty}, \,R^*=\frac{R}{R_0},\,  r^*=\frac{r}{R_0},\, t^*=\frac{D_s t}{R_0^2}.
\end{equation}
Substituting in Eq.~\eqref{eq:transport} and dropping stars in what follows for notational convenience,  we obtain the colloid dynamics as solution to
\begin{equation}
\frac{dR}{dt}=\alpha_1\frac{\partial c}{\partial r}\bigg\rvert_{r=R}
\end{equation}
with $c$ solution to
\begin{equation}\label{eq:3soladvdiff}
\frac{\partial c}{\partial t} + \frac{R^2}{r^2}(\alpha_1-\beta_1)\frac{\partial c}{\partial r}\bigg\rvert_{r=R}\frac{\partial c}{\partial r}= \left(\frac{\partial^2 c}{\partial r^2}+\frac{2}{r}\frac{\partial c}{\partial r}\right),
\end{equation}
with dimensionless boundary conditions
\begin{equation}
c(R(t),t)=1,\,\, t\geq 0,
\quad{\rm and}\quad
c(r,0)=0,\,\, r>1,
\end{equation}
where we have defined the two dimensionless parameters
\begin{equation}
\alpha_1=\frac{c_0-c_\infty}{\rho_s},\quad \beta_1=\frac{c_0-c_\infty}{\rho_l}\cdot
\end{equation}
We note that despite a negligibly small solute P\'eclet number, it was necessary to include an advective term due to volume conservation, whose relative strength is given by $(\alpha_1-\beta_1)$. It is therefore independent of the P\'eclet number and its irrelevance at leading order will be  only a consequence of the weak solubility assumption. Only when there is no density mismatch between colloid and fluid  is this term identically  zero. Furthermore, the swimmer radius remains constant when the solvent is saturated with solute, as may be expected intuitively.
 
\subsubsection{Asymptotic solution}\label{sec:asympsol}
In order to make analytical progress, we make the assumptions that
\begin{equation}
\alpha_1,\beta_1 \ll 1,
\end{equation}
which corresponds to a {low-solubility} limit for the colloid material. We can then develop an asymptotic expansion to solve for $c$ and $R$. Here we will only calculate the leading-order solution, but our setup allows for calculations to arbitrarily high orders. We proceed by a rescaling of our spatial coordinate as
\begin{equation}
x=\frac{r}{R},\quad y(x,t)=xc(x,t),
\end{equation}
so that our system becomes
\begin{align}
R^2\frac{\partial y}{\partial t}&+R\dot{R}y+(\alpha_1-\beta_1)\left(\frac{1}{x^2}\frac{\partial y}{\partial x}-\frac{y}{x^3}\right)\left(\frac{\partial y}{\partial x}\bigg\rvert_{x=1}-1\right)\nonumber\\
&=\frac{\partial^2 y}{\partial x^2}
\end{align}
and
\begin{equation}\label{eq:3rsquared}
R^2=1+2\alpha_1\left(\int_{0}^{t}\frac{\partial y}{\partial x}\bigg\rvert_{x=1}dt' -t\right)
\end{equation}
with boundary conditions
\begin{equation}
y(1,t)=1,\quad y(x,0)=0.
\end{equation}
The solution may be written as
\begin{equation}
y(x,t;\alpha_1, \beta_1)=y_0(x,t)+\alpha_1 y_\alpha(x,t) + \beta_1 y_\beta(x,t) +o(\alpha_1,\beta_1).
\end{equation}
The problem for $y_0$ reduces to the one-dimensional heat equation with Dirichlet boundary conditions and its solution is well known to be
\begin{equation}
y_0(x,t) = \text{erfc}\left(\frac{x-1}{2\sqrt{t}}\right),
\end{equation}
whence to leading order
\begin{equation}
R^2=1-2\alpha_1\left(t+2\sqrt{\frac{t}{\pi}}\right),
\end{equation}
or, after reinserting dimensions, we obtain   our desired result
\begin{equation}\label{eq:nrradius}
R(t) = R_0\sqrt{1-2\alpha_1\left(\frac{t}{t_s}+\frac{2}{\sqrt{\pi}}\sqrt{\frac{t}{t_s}}\right)}.
\end{equation}
where $t_s=R_0^2/D_s$ is the diffusive time scale for the solute. An illustration of this decay, along with a comparison to the reacting model is presented in Fig.~\ref{fig:rcomp}. Denoting by $T_d$ the finite time at which the particle disappears, and taking into account the order of terms we neglect, we can deduce that
\begin{equation}\label{eq:timeofdeath}
\begin{split}
T_d&=\frac{t_s}{2\alpha_1}\left(1-\sqrt{\frac{8}{\pi}}\sqrt{\alpha_1}+\mathcal{O}(\alpha_1,\beta_1)\right).
\end{split}
\end{equation}
Therefore at leading order, the lifetime of the colloid scales inversely proportional with the solubility and diffusivity of its material, but quadratically with the initial colloid radius $R_0$. However, the correction from the next-to-leading order term remains significant for $\alpha_1\gtrsim10^{-3}$ due to its slow square-root like decay.

\subsubsection{Physical interpretation}\label{sec:physint}
The aim of this section is to provide some physical interpretation for Eq.~\eqref{eq:nrradius}. 
For many ions in water at room temperature, the diffusivity is approximately $D_s\sim10^{-9}$~m$^2$/s \cite{haynes2014crc}. In the case of an initially micron-sized colloid this gives
\begin{equation}
t_s\sim 10^{-3}\,{\rm s}.
\end{equation}
The other (previously unknown) time scale in the problem is the swimmer lifetime $T_d$. There is a separation of scales that is to leading order inversely proportional to $\alpha_1$. In the specific example of calcium carbonate with $\alpha_1\approx10^{-6}$ \cite{haynes2014crc}, we obtain
\begin{equation}
T_d\sim 10^{3}\,{\rm s}\sim 10 \text{ min},
\end{equation}
which is a conceivably desirable lifetime for a microswimmer. 

The separation of scales has further consequences for the decay rate. For $t\ll t_s$ we have $R^2 \sim 1-4\sqrt{t\alpha_1^2/t_s\pi}$, while for $t\gg t_s$ we obtain the behaviour $R^2\sim1-2\alpha_1 t/t_s$. Therefore the particle size satisfies $R\sim\sqrt{1-2\alpha_1 t/t_s}$ except for a short, transient period on the order of $t_s$. This feature may be explained physically. Initially, the discontinuity in concentration at $r=R$ causes a large concentration gradient and fast dissolution but on the (fast) scale of solute diffusion the system relaxes to equilibrium in a boundary layer of thickness $\sim\sqrt{D_s t_s}$, which is on the order of the colloid size, $R_0$. From this point onwards the colloid is surrounded by a cloud of solute in equilibrium and the process becomes quasi-static. At leading order, the dissolution dynamics therefore reduces to steady diffusion. This gives simultaneously justification to our assumption of sphericity, since the diffusive boundary layer smooths out any surface inhomogeneities. 

As an aside, we note while the dissolution process of microbubbles is driven by capillary pressures~\cite{michelin2018collective}, the $R\sim \sqrt{1-t}$ behaviour also emerges in the absence of surface tension, essentially also due to the dominance of diffusive effects.

Finally, we point out that $\alpha_1$ and $\beta_1$ depend only on the material chosen for the swimmer (and its abundance in the bulk fluid). Unsurprisingly, only materials that are considered insoluble on the macroscale yield appreciable microswimmer life times. Hence,  together with fine tuning of the initial radius $R_0$, full control of the dissolution dynamics can be achieved through the microswimmer design.

\subsection{Dissolution through reaction}

\begin{figure}[t!]
	\centering
	\includegraphics[width=0.5\columnwidth]{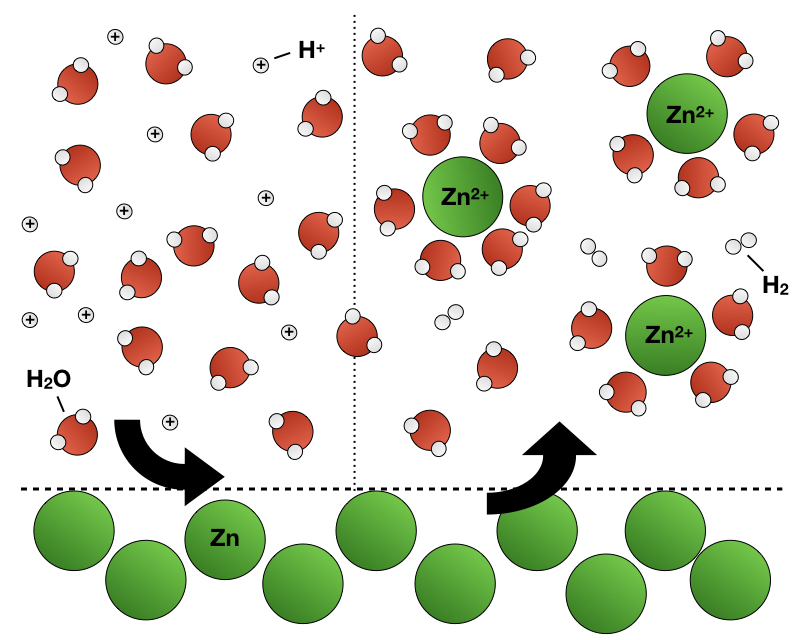}
	\caption{Schematic presentation  of the molecular dynamics near the boundary of a reacting colloid. In this example, motivated by experiments in Ref.~\cite{chen2016transient}, zinc is dissolved in acid forming zinc-ions and molecular hydrogen. If $\text{Da}=0$, i.e.~infinitely much H$^+$ is present to sustain the reaction, the dissolution rate is constant. If $\text{Da}>0$, the reaction rate will depend on the amount of fuel present, but not on the amount of product. }\label{fig:znreaction}
\end{figure}

Artificial microswimmers are rarely composed solely  of chemically inert materials. Indeed, autophoretic swimmers often consume a fuel in the solvent, like in the widely studied case of catalytic platinum swimmers splitting hydrogen peroxide into water and oxygen \cite{moran2017phoretic}. A sketch of the process is illustrated in Fig.~\ref{fig:znreaction} in the specific case of zinc dissolving in acid as realised experimentally by Chen et.\ al.\ \cite{chen2016transient}. An analogous picture may be imagined for the case of biodegradation by enzymes.

A degradable autophoretic colloid might therefore consist of a reactant that will then dissolve into the fluid. To this end, let us consider a fixed reaction-rate boundary condition. It will be important to distinguish between the concentration of fuel $c_f(\r,t)$  and concentration of swimmer substrate $c_s(\r,t)$. For example, in the case of zinc, the fuel concentration might be provided by hydrogen ions in acid, which relates their concentration directly to the pH-value of the solvent, while  the concentration of substrate  influences the dissolution rate through mass conservation. Notation-wise, we will use  the subscript $f$ to refer below to the fuel and the subscript $s$ to the substrate. 

\subsubsection{Mathematical model}
The mathematical development is similar to the non-reacting swimmer, with an important change to the boundary conditions. Indeed, unlike Eq.~\eqref{eq:3solbc} where the concentration at the boundary was fixed, 
the boundary conditions for the  fields $c_s$ and $c_f$ are now given by
\begin{equation}\label{eq:28}
-D_s{\n}\cdot\nabla c_s|_R = k_s c_f,\quad -D_f{\n}\cdot\nabla c_f|_R = -k_f c_f,
\end{equation}
where $k_s$ and $k_f$ are the constant reaction rates for solute and fuel respectively and $D_f$ the diffusivity of fuel in the solution. Mass conservation for the colloidal particle leads to
\begin{equation}\label{eq:Rcf}
\frac{dR}{dt}=-\frac{k_s c_f(R)}{\rho_s}.
\end{equation}
Furthermore, we once again have conservation of fluid volume giving rise to a source flow
\begin{equation}
\u=\dot{R}(1-\rho_s/\rho_l)\frac{R^2}{r^2}{\hat{\r}}.
\end{equation}
Similar to what was done above, we assume that the P\'eclet numbers associated with the solute and the fuel dynamics are small, so that only volume conservation gives rise to advective flows. We can then write the advection-diffusion equation for $c_f$ as
\begin{equation}\label{eq:cf}
\frac{\partial c_f}{\partial t} - (1-\rho_s/\rho_l)\frac{k_sc_f(R)}{\rho_s}\frac{R^2}{r^2}\frac{\partial c_f}{\partial r}= D_f \left(\frac{\partial^2 c_f}{\partial r^2}+\frac{2}{r}\frac{\partial c_f}{\partial r}\right).
\end{equation}
Introducing non-dimensionalised variables as
\begin{equation}
c_f^*=\frac{c_f}{c_{f,\infty}}, R^*=\frac{R}{R_0}, r^*=\frac{r}{R_0},t^*=\frac{D_f t}{R_0^2},
\end{equation}
where  $c_{f,\infty}$ is the mass concentration of fuel in the bulk, 
we may substitute in Eqs.~\eqref{eq:Rcf} and \eqref{eq:cf} and dropping stars immediately we find
\begin{equation}
\frac{\partial c_f}{\partial t} - \text{Da}(\alpha_2-\beta_2)c_f(R)\frac{R^2}{r^2}\frac{\partial c}{\partial r}= \frac{\partial^2 c}{\partial r^2}+\frac{2}{r}\frac{\partial c}{\partial r},
\end{equation}
\begin{equation}
\frac{dR}{dt}=-\text{Da}\alpha_2 c_f(R),
\end{equation}
with the boundary conditions
\begin{equation}
\begin{aligned}
&c_f\to 1,\quad&r\to\infty,\\
&\frac{\partial c_f}{\partial r}=\text{Da} c_f,\quad&r=1,\\
&c_f(r,0)=1,\quad&r>1,\\
&R(0)=1,&
\end{aligned}
\end{equation}
where we have defined the three dimensionless numbers
\begin{equation}
\text{Da}=\frac{R_0 k_f}{D_f},\quad \alpha_2=\frac{c_{f,\infty}k_s}{\rho_s k_f},\quad \beta_2=\alpha_2\frac{\rho_s}{\rho_l}\cdot
\end{equation}
Here $\text{Da}$ is a Damk\"ohler number for the fuel, indicating the ratio between reactive and diffusive fluxes, while $\alpha_2$ and $\beta_2$ may be interpreted as dimensionless ratios comparing the mass of fuel consumed against the mass of solute shed in the reaction. 

 Upon rescaling our coordinates according to
\begin{equation}
x=\frac{r}{R},\quad y(x,t)=c_f x,
\end{equation}
our system becomes
\begin{equation}
R^2\frac{\partial y}{\partial t}+R\dot{R}y-\text{Da}(\alpha_2-\beta_2)R \left(\frac{1}{x^2}\frac{\partial y}{\partial x}-\frac{y}{x^3}\right)y(1,t)=\frac{\partial^2 y}{\partial x^2},
\end{equation}
and
\begin{equation}
R=1-\text{Da}\alpha_2\int_{0}^{t}y(1,t')dt',
\end{equation}
with
\begin{equation}\label{eq:robinbc}
y(x,0)=1,\quad \frac{\partial y}{\partial x}(1,t)=\text{Da} y(1,t).
\end{equation}
From here, we can again proceed by means of an asymptotic expansion.

\subsubsection{Asymptotic expansion}
We next assume $\alpha_2 \text{Da},\,\beta_2 \text{Da} \ll 1$ and write the solution as a power expansion
\begin{align}
&y(x,t;\alpha_2, \beta_2;\text{Da})=\nonumber\\
&y_0(x,t;\text{Da})+\alpha_2 y_\alpha(x,t;\text{Da}) + \beta_2 y_\beta(x,t;\text{Da}) +{\rm h.o.t.}
\end{align}
The boundary condition in Eq.~\eqref{eq:robinbc} consitutes a Robin problem and can be solved by considering the quantity $\phi=y-\text{Da}^{-1} \partial y/\partial x$ subject to Cauchy conditions \cite{carslaw1959heat}. The solution for $y_0$ is
\begin{align}
y_0(x,t;\text{Da})=&\text{erf}\left(\frac{x-1}{2\sqrt{t}}\right)\nonumber\\
&+e^{\text{Da}(x-1)+\text{Da}^2 t}\text{erfc}\left(\frac{x-1}{2\sqrt{t}}+\text{Da}\sqrt{t}\right).
\end{align}
It follows that
\begin{equation}
y_0(1,t;\Da) =e^{\text{Da}^2 t}\text{erfc}\left(\text{Da}\sqrt{t}\right),
\end{equation}
and hence to leading order in $\alpha_2$,
\begin{equation}
R(t)=1-2\alpha_2\sqrt{\frac{t}{\pi}}-\frac{\alpha_2}{\text{Da}}\left[e^{\text{Da}^2 t}\text{erfc}\left(\text{Da}\sqrt{t}\right)-1\right].
\end{equation}
Upon reinserting dimensions we finally arrive at
\begin{align}\label{eq:model3diss}
R(t)=R_0\Bigg\{&1-\alpha_2\frac{2}{\sqrt{\pi}}\sqrt{\frac{t}{t_f}}\nonumber\\
&\left. -\frac{\alpha_2}{\text{Da}}\left[e^{\text{Da}^2 t/t_f}\text{erfc}\left(\text{Da}\sqrt{\frac{t}{t_f}}\right)-1\right]\right\}.
\end{align}
where $t_f=R_0^2/D_f$ is the diffusive time scale for the fuel. 

\subsubsection{Slow reaction limit (fixed solute flux)}

Inspired by a  study of boundary conditions in the context of finite P\'eclet-number propulsion in Ref.~\cite{michelin2014phoretic}, we may consider separately the limits $\text{Da}\to0$ and $\text{Da}\to\infty$. 
Each of these limits will lead to a different model that we will consider in the remainder of this paper. 

For small Damk\"ohler number,  we find
\begin{equation}\label{eq:srradius}
\begin{aligned}
&R(t) = R_0\left[1- \frac{\alpha_2 \text{Da}}{t_f} t +\mathcal{O}\left( \text{Da}^2{\left(\frac{t}{t_f}\right)}^{3/2}\right)\right],\nonumber\\
&\text{as Da}\to 0,\quad\frac{t}{t_f}\lesssim \text{Da}^{-2}.
\end{aligned}
\end{equation}
When $\text{Da}=0$, no reaction takes place and the radius of the colloid  remains constant. At next to leading order we have linear decay, so the lifetime $T_d$ is
\begin{equation}
T_d=\frac{t_f}{\alpha_2}\text{Da}^{-1} =\frac{R_0\rho_s}{c_{f,\infty}k_s} \quad (\text{Da}\to0),
\end{equation}
which is consistent with the asymptotic expansion to this order. Thus we arrive at a model for the dissolution with a constant solute flux. We note the different scaling compared to the non-reacting model where the lifetime scaled as $T_d\sim R_0^2$. This is indicative of the absence of diffusion in this limit. Note that the model can be recovered from simply applying mass conservation to a flux boundary condition of the form
\begin{equation}
-D_s\frac{\partial c}{\partial r}\bigg\vert_{r=R(t)}=c_{f,\infty}k_s,
\end{equation}
which shows that the flux is equal to $c_{f,\infty}k_s$.

\subsubsection{Fast reaction limit}
Conversely, as $\text{Da}\to\infty$  (still with $\alpha_2 \text{Da}\ll 1$), we find that
\begin{equation}\label{eq:frradius}
\begin{aligned}
&R(t) = R_0\left\{1-\frac{2\alpha_2}{\sqrt{\pi}}\sqrt{\frac{t}{t_f}} +\mathcal{O}\left(\text{Da}^{-1}\right) \right\},\nonumber\\
&\text{as Da}\to \infty, \quad \frac{t}{t_f}\gtrsim \text{Da}^{-2}.
\end{aligned}
\end{equation}
In this limit the reaction is infinitely fast, so the boundary condition on the fuel effectively reduces to instantaneous depletion, $c_f(R,t)=0$, and the dissolution rate is limited by the diffusive flux of fuel from the bulk. Correspondingly the lifetime $T_d$  in dimensional units is 
\begin{equation}
T_d=\frac{\pi}{4\alpha_2^2}\frac{R_0^2}{D_f}\quad (\text{Da}\to\infty,\alpha_2 \text{Da}\ll 1),
\end{equation}
a result which is again consistent with the expansion. Apart from the introduction of reaction rates, this result is qualitatively different from the non-reacting swimmer insofar as the lifetime depends on the square of swimmer density and reactant concentration at infinity, rather than being inversely proportional to solubility. We remark that in the case of hydrogen ions, the concentration $c_f$ is directly related to the pH value of the solvent, which establishes an experimentally accessible relationship between the pH and swimmer dissolution dynamics.

\begin{figure}[t!]
	\centering
	\includegraphics[width=0.5\columnwidth]{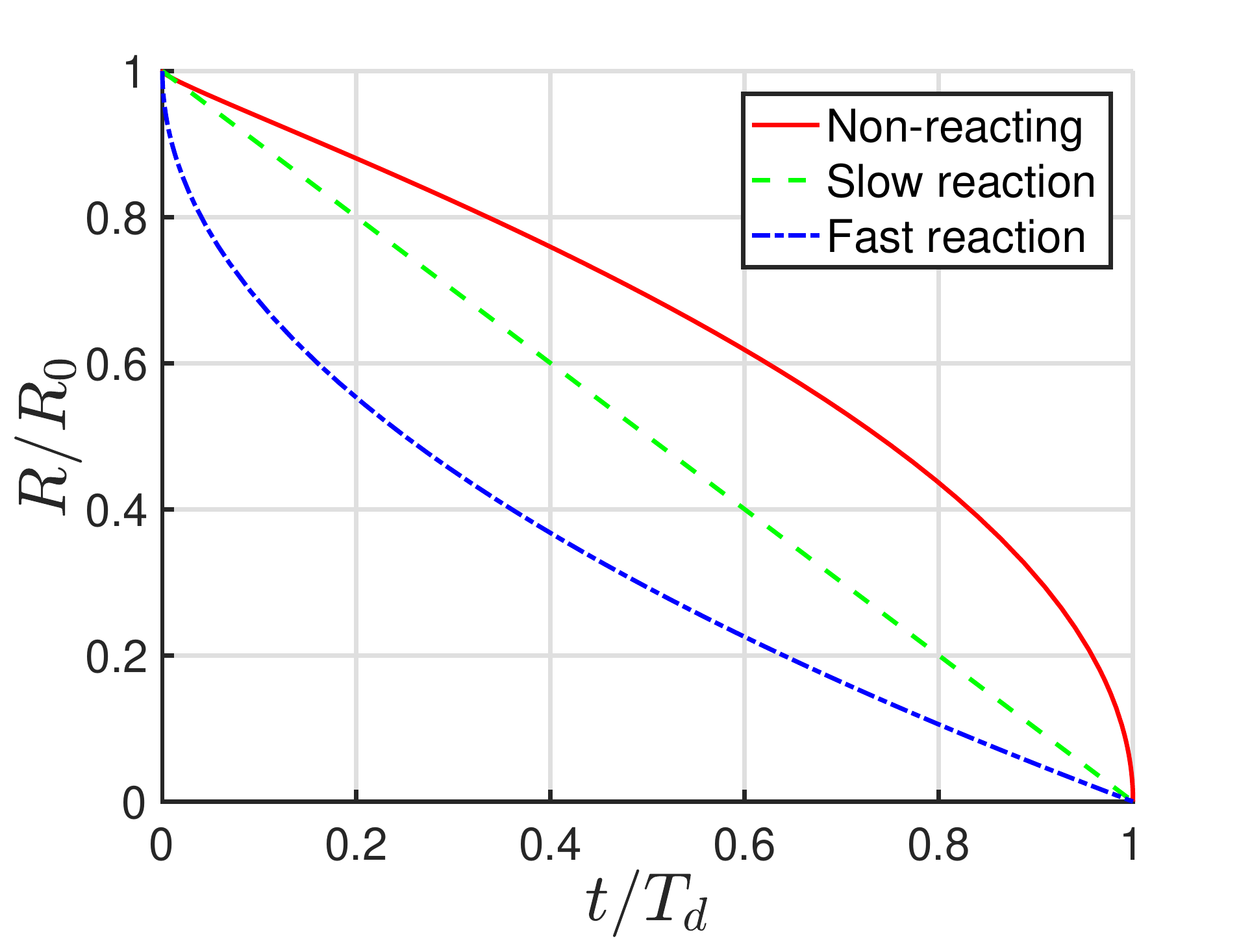}
	\caption{Comparison of the decay dynamics between the three models: decay of the dimensionless colloid radius as a function of dimensionless time.	(i) Non-reacting (red solid line; $t_d/T_d=0.01$); 
(ii) Slow reaction (green dashed line);
(iii) Fast reaction (blue dash-dotted line). 
 }\label{fig:rcomp}
\end{figure}

In Fig.~\ref{fig:rcomp} we illustrate the different decay behaviour for our three models: 
(i) Non-reacting (red solid line, with $t_d/T_d=0.01$); 
(ii) Slow reaction (green dashed line);
and (iii) Fast reaction (blue dash-dotted line). 
 We note for the non-reacting model the decay rate increases with time, whereas it is constant for the slowly reacting, and decreasing for the fast reacting model. In the following two sections, we will explore the important consequences this has for the stochastic behaviour of dissolving microswimmers.

\section{Passive dynamics of dissolving colloids}\label{sec:Passive}

After developing three models for the dissolution of a spherical colloid, we now ask what effect this reduction in size has on its fluctuating trajectory. As will be shown, the mean squared displacement of a stochastic self-propelled particle is given by the sum of the contributions from translational noise and active motion. This allows us to split the analysis into the case of a passive colloid with no intrinsic propulsion mechanism but with translational noise and an active colloid with rotational but no translational diffusion. We treat the former case in this section and consider the motion of self-propelled particles in \S\ref{sec:Active}.

\subsection{Mathematical model}
The  change in the dynamics of colloidal particles arises through the time dependence of the translational diffusion coefficient, which is given by the Stokes-Einstein relation \cite{einstein1905motion}
\begin{equation}
D(t)=\frac{k_B T}{6\pi\mu R(t)}\equiv D_0\frac{R_0}{R(t)},
\end{equation}
where $k_B$ is Boltzmann's constant, $T$ is absolute temperature  and $D_0\equiv D(0)=k_B T/6\pi\mu R_0$. In analogy with classical Brownian motion, we consider the following overdamped Langevin equation for the  position of the passive colloidal particle, $\r(t)$,
\begin{equation}\label{eq:pasLangevin}
d\r=\sqrt{2D(t)}d\W.
\end{equation}
Classically, $\W(t)$ is white noise with the properties that
\begin{equation}\label{eq:whitenoiseprop}
	\langle d\W \rangle ={\bf{0}},\quad \langle dW_i(t)dW_j(t')\rangle=\delta_{ij}\delta(t-t')dt,
\end{equation}
with brackets denoting ensemble averages. The right-hand side of Eq.~\eqref{eq:pasLangevin} therefore varies on two different time scales: the rate of change of $D$ and the time scale of the molecular chaos $\tau_{MC}$ that gives rise to noise. Typically, $\tau_{MC}=\mathcal{O}(10^{-13}s)$ \cite{haynes2014crc}. The mathematical assumption of $\delta$-correlated noise only holds true if $\tau_{MC}$ is very small compared to the time scale of diffusion, which holds true for microscopic colloids. However, since the rate of change of $D$ diverges as the swimmer size tends to 0, this model is expected break down at the very end of the swimmer lifetime. In the case of the non-reacting model this singularity is integrable and poses no problem, whereas for the reacting model we will also include a physical discussion of the breakdown. 

For an active self-propelled particle at velocity ${\bf U}(t)$,  the right-hand side of the Langevin equation Eq.~\eqref{eq:pasLangevin} includes an additional term ${\bf U}(t)dt$, which is deterministic in the sense that it is uncorrelated with translational white noise (even if ${\bf U}(t)$ is subject to rotational noise). A straightforward integration using the properties in Eq.~\eqref{eq:whitenoiseprop} then shows that the total mean squared displacement is given by the sum of active and passive contributions, 
\begin{equation}
	\langle r^2 \rangle_{tot} = 	\langle r^2 \rangle_a + 	\langle r^2 \rangle_p,
\end{equation}
as claimed.

The stochastic dynamics in Eq.~\eqref{eq:pasLangevin} gives rise to a Fokker-Planck equation for the probability  for the position of the particle, $P(\r,t)$, as
\begin{equation}
\frac{\partial P}{\partial t}= D(t) \nabla^2 P.
\end{equation}
We can solve this by a rescaling of time, introducing $\tau(t)$ such that
\begin{equation}\label{eq:3deftau}
\tau = \int_{0}^{t} D(s) ds=D_0\int_{0}^{t}\frac{R_0}{R(s)}ds,
\end{equation}
which yields
\begin{equation}
\frac{\partial \tilde P}{\partial \tau}= \nabla^2 \tilde P.
\end{equation}
where $\tilde{P}(\r,\tau)=P(\r,t)$. In three spatial dimensions this equation has a well known Gaussian solution corresponding to the initial condition of a particle located at the origin,
\begin{equation}
\tilde P(\r,\tau)=\tilde P(r=|\r|,\tau)=\frac{1}{(4\pi\tau)^{3/2}}\exp\left(-\frac{r^2}{4\tau}\right).
\end{equation}
The first two moments are well known to be $\langle \r\rangle = {\bf 0}$ and $\langle {r}^2\rangle = 6\tau$. The total passive mean squared displacement of the particle in its lifetime, $\langle r^2 \rangle_p\equiv\langle r^2 \rangle(T_d)$, is therefore given by the integral
\begin{equation}\label{eq:pasmsd}
	\langle r^2 \rangle_p=6D_0\int_{0}^{T_d}\frac{R_0}{R(t)}dt.
\end{equation}
Note that since $R\leq R_0$, the integral has value larger than $T_d$. Therefore dissolution always enhances passive diffusion. All that remains to be done is to calculate the integral for each of our three models.

\subsection{Total root mean squared displacement}
In the following we consider the solutions to Eq.~\eqref{eq:pasmsd}. Bearing in mind the order of terms we neglected in the derivation of Eq.~\eqref{eq:nrradius}, we can integrate Eq.~\eqref{eq:pasmsd} directly to obtain the following result for the non-reacting model
\begin{equation}
\langle {r}^2\rangle_{p} = 6D_0\times\frac{t_s}{\alpha_1}\left(1-\sqrt{\frac{\pi}{2}}\sqrt{\alpha_1}+\mathcal{O}(\alpha_1,\beta_1)\right).
\end{equation}
Comparing with Eq.~\eqref{eq:timeofdeath} we can see that at leading order in $\alpha_1$, dissolution enhances the total mean squared displacement by a factor of two. Through the scaling of $t_s$ with $R_0$ we also find that $\langle {r}^2\rangle_p\sim R_0$. This may be tested easily in experiments without affecting the other parameters. Perhaps surprisingly, this also means that in contrast to fixed-size swimmers, the importance of passive Brownian effects increases with swimmer size, since the smaller diffusivity is overcompensated for by the longer life span. The scaling with $\alpha_1$ can be explained the same way, as a colloid with small $\alpha_1$ decays slower, lives longer and therefore travels further.

For the slow reaction model we can use Eq.~\eqref{eq:srradius} in the integration of Eq.~\eqref{eq:pasmsd} to find
\begin{equation}
\langle r^2\rangle(t) =6D_0\times T_d\log\left(\frac{R_0}{R(t)}\right).
\end{equation}
This expression diverges logarithmically  as $t\to T_d$. This should not be taken as indicative of superdiffusion, but can be resolved by the breakdown of the Stokes-Einstein relation below a certain colloid size. Past experiments suggest this happens for colloids smaller than a few nanometres in diameter \cite{li2009critical}. Compared to an initial colloid size on the scale of a few microns, this corresponds to 2 to 4 orders of magnitude. Since the divergence of the mean squared displacement is logarithmic, this will give a total mean squared displacement that is greater than that of a non-dissolving colloid by a factor of $\mathcal{O}(1)-\mathcal{O}(10)$. Furthermore, since $D_0T_d$ is independent of $R_0$ for this model, the contribution of passive Brownian motion only depends weakly on the initial colloid size. This is in contrast with the other models, and indicative of the absence of diffusion. 

Finally, using Eq.~\eqref{eq:frradius} in Eq.~\eqref{eq:pasmsd} we obtain  for the fast reaction limit the result
\begin{equation}
\langle {r}^2 \rangle (t) = 6D_0\times 2T_d\left(\log\left(\frac{R_0}{R(t)}\right)+\frac{R(t)}{R_0}-1\right).
\end{equation}
where again we have a logarithmic divergence as $t\to T_d$. Using previous definitions we find that as in the non-reacting model $\langle {r}^2 \rangle_p\sim R_0$ (+ logarithmic corrections) and also that $\langle {r}^2 \rangle_p\sim \alpha_2^{-2}$. The passive mean squared displacement therefore depends rather sensitively on the availability of fuel for the reaction.

\section{Active motion of dissolving colloids}\label{sec:Active}

After examining the dynamics of passive particles, we now turn to the effect of dissolution on self-propelled microswimmers. For the case of active particles subject to rotational diffusion with coefficient $D_r$, it is well known that self-propulsion at velocity $U$ gives rise to an effective enhanced translational diffusivity \cite{golestanian2007designing}
\begin{equation}
	D_{\text{eff}}=D+\frac{U^2}{6D_r},
\end{equation}
for times much longer than $D_r^{-1}$, the time scale of rotational diffusion (i.e.~in the limit $t D_r\gg1$). On scales much shorter than this the motion is instead ballistic, i.e.~$\langle r^2 \rangle \sim U^2 t^2$.

\begin{figure}
	\centering
	\includegraphics[width=0.5\columnwidth]{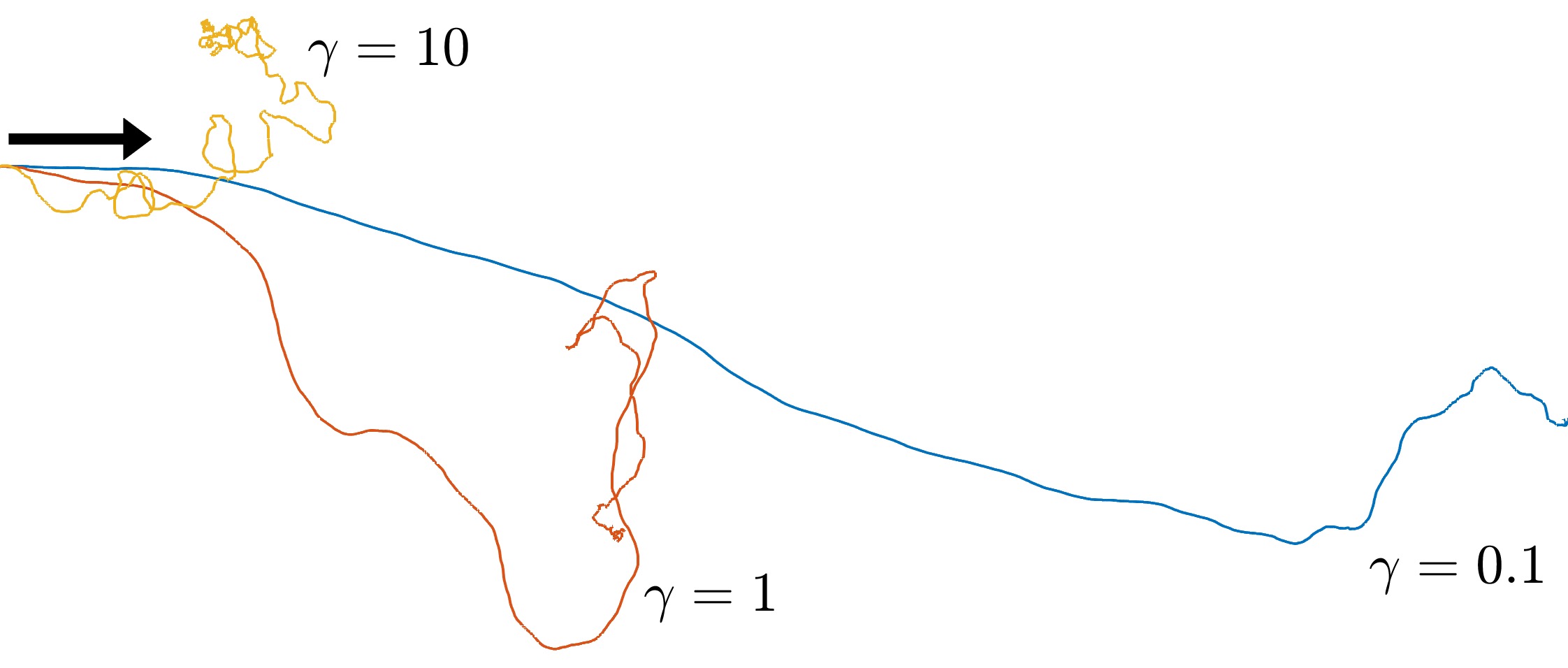}
	\caption{2D-projections of sample trajectories for different values of $\gamma$. The colloids initially swim from left to right (see arrow) and dissolve according to the non-reacting model with the same length scale and lifetime.}\label{fig:illustration}
\end{figure}

In this new scenario however, an additional scale is introduced through the swimmer lifetime, $T_d$. It is therefore vital to consider the dimensionless quantity
\begin{equation}
	\gamma := D_{r,0}T_d,
\end{equation}
where we define $D_{r,0}=k_BT/8\pi\mu R_0^3$. If $\gamma\lesssim 1$, then the particle disappears before displaying macroscopically diffusive behaviour. Conversely, if $\gamma\gtrsim 1$ we expect trajectories that are qualitatively similar to that of a classically diffusive colloid at long time scales. The qualitative role of $\gamma$ is illustrated in Fig.~\ref{fig:illustration} where we observe  three trajectories becoming more curly as time progresses, since diffusivity increases as the swimmer dissolves. However, only colloids with large values of $\gamma$ (here, $\gamma=10$) exist long enough for this effect to become significant, giving rise to a macroscopically `diffusive' trajectory. Conversely, for small $\gamma$ (here, $\gamma=0.1$) trajectories appear macroscopically `ballistic'. Depending on the application, it may be desirable to design swimmers that belong to either of these two regimes. In water at room temperature we have $D_{r,0}^{-1}\approx 6(R_0/\mu \text{m})^3 \text{ s}$ \cite{haynes2014crc}, so depending on the initial colloid size the threshold lifetime ranges from seconds to hours. Therefore both regimes are conceivable for applications and thus relevant to study. We proceed with the development of our theoretical framework to derive expressions for the active mean squared displacement and present analytical solutions for each model both as $\gamma\to 0$ and as $\gamma \to \infty$. We then validate our theoretical results against numerical simulations of the associated Langevin dynamics.

\subsection{Mathematical model}\label{sec:act_matmod}
In the rest of this section we assume that the colloid is subject to Langevin dynamics as
\begin{align}\label{eq:actLangevin}
	d\r &=U\e dt, \\
	d\e&=-2D_r(t)\e dt+\sqrt{2D_r(t)}\bPi(\e)\cdot d\W,
\end{align}
to be understood in the It\^o formulation of stochastic calculus. Here $U$ is the particle self-propulsion speed, $\e$ the unit vector along the direction of propul d $\Pi_{ij}=\delta_{ij}-e_ie_j$. As is the case for a wide range of phoretic swimmers \cite{moran2017phoretic}, we assume the velocity $U$ to be independent of the swimmer size. Moreover, we set $D=0$ to isolate the effect of active diffusion, which generally exceeds that of (regularised) passive diffusion discussed previously. Since both contribute independently however, they may simply be added together if the total mean squared displacement is desired. We also neglect the details of the propulsion mechanism and possible interactions with our dissolution models.

As in the classical case, the $\e$-dynamics decouple from the $\r$-dynamics. With the same assumptions regarding the separation of time scales as in the passive case, $\e(\theta,\phi)$ is therefore subject to the Fokker-Planck equation
\begin{equation}
	\frac{\partial}{\partial t}P(\theta,\phi,t)=D_r(t)\nabla_{\text{ang}}^2P,
\end{equation}
where $\nabla_{\text{ang}}^2$ denotes the angular part of the Laplacian operator. By introducing a rescaled time $\tau_r(t)$ as
\begin{equation}
	\tau_r=\int_0^t D_r(s)ds=D_{r,0}\int_0^t\left(\frac{R_0}{R(s)}\right)^3ds,
\end{equation}
this may be used to show that $\langle \e(t)\cdot\e(0)\rangle=\exp(-2\tau_r)$. Therefore we have the following expression for the total active mean squared displacement,
\begin{equation}\label{eq:actmsd}
	\langle r^2 \rangle_a = 2U^2\int_{0}^{T_d} dt' \int_{0}^{t'}dt'' \exp\bigg\{{-2\left[\tau_r(t')-\tau_r(t'')\right]} \bigg\}.
\end{equation}
Substituting values for our models and rescaling variables, this gives the following general expressions.
\begin{align}
\langle r^2 \rangle_a&=U^2T_d^2\int_{0}^{\infty}dx'\int_{0}^{x'}dx''\frac{2e^{-2\gamma(x'-x'')}}{(1+x'/2)^3(1+x''/2)^3},\nonumber\\
&\text{(non-reacting)}\label{eq:amsd1}\\
\langle r^2 \rangle_{a}&= 	U^2T_d^2\int_{0}^{\infty}dx'\int_{0}^{x'}dx''\frac{2e^{-2\gamma(x'-x'')}}{(1+2x')^{3/2}(1+2x'')^{3/2}},\nonumber\\
&\text{(slow reaction)}\label{eq:amsd2}\\
\langle r^2 \rangle_{a}&=U^2T_d^2\int_{0}^{\infty}dx'\int_{0}^{x'}dx''\frac{2e^{-2\gamma(x'-x'')}}{(1+\sqrt{x'})^{3}(1+\sqrt{x''})^{3}}.\nonumber\\
&\text{(fast reaction)}\label{eq:amsd3}
\end{align}
Unfortunately, while these are exact results,  it is not possible to evaluate these integrals analytically for arbitrary values of $\gamma$. However, we can derive asymptotic solutions in both the diffusive and ballistic limits, as we now show.

\subsubsection{Diffusive limit ($\gamma\to\infty$)}\label{sec:actdiff}
In the diffusive limit, $\gamma\gg 1$,  we can use Watson's lemma to develop an asymptotic expansion, with details given in the Appendix. In the case of a non-reacting swimmer,  we find
\begin{equation}\label{eq:actmsd1}
	\langle r^2 \rangle_a\sim\frac{2}{5}\frac{U^2T_d}{D_{r,0}}\left[1-\frac{5}{8\gamma}+\dots\right],\quad\gamma\to\infty\quad\text{(non-react.)}.
\end{equation}
As expected, the behaviour is diffusive  and the leading-order scaling is
\begin{equation}
\langle r^2 \rangle_a\sim\frac{U^2\mu\rho_s R_0^5}{k_BT D_s(c_0-c_\infty)},\quad\gamma\to\infty\quad\text{(non-reacting)}.
\end{equation}
We notice the appearance of the $2/5$ factor in Eq.~\eqref{eq:actmsd1}, indicating that the enhancement of the diffusivity through active motion is reduced dramatically, to just 40\% of that of a comparable classical colloid. Furthermore, the active mean squared displacement scales as $\sim R_0^5$, making the range of the swimmer extremely sensitive to its initial size. This scaling breaks down for very large swimmers, since it is necessary that $\gamma\sim R_0^{-1}$ is sufficiently large for this expansion to remain valid.

For the slowly reacting swimmer we find in a similar fashion that
\begin{equation}\label{eq:actmsd2}
\langle r^2 \rangle_a\sim\frac{1}{4}\frac{U^2T_d}{D_{r,0}}\left[1-\frac{1}{\gamma}+\dots\right],\quad\gamma\to\infty\quad\text{(slow react.)}.
\end{equation}
with the leading-order scaling
\begin{equation}
\langle r^2 \rangle_a\sim\frac{U^2\mu\rho_s R_0^4}{k_BT c_{f,\infty}k_s},\quad\gamma\to\infty\quad\text{(slow reaction)}.
\end{equation}
We see that the diffusivity in Eq.~\eqref{eq:actmsd2} is reduced even further, to 25\% that of a classical colloid. Finally for the fast reacting swimmer we obtain
\begin{equation}\label{eq:actmsd3}
\langle r^2 \rangle_a\sim\frac{1}{10}\frac{U^2T_d}{D_{r,0}}\left[1-\frac{5}{2\gamma}+\dots\right],\quad\gamma\to\infty\quad\text{(fast react.)/},
\end{equation}
and the leading-order scaling
\begin{equation}
\langle r^2 \rangle_a\sim \frac{U^2\mu \rho_s^2k_f^2 R_0^5 }{k_B T  D_fc_{f,\infty}^2 k_s^2},\quad\gamma\to\infty\quad (\text{fast reaction}).
\end{equation}
This third dissolution model gives the strongest reduction of the active mean squared displacement in the diffusive regime, to just 10\% that of a classical colloid.

The strong reduction in mean squared displacement across all three models suggests that it is impractical to rely on active diffusion to transport dissolving microswimmers. Instead designs may be aimed at exploiting the ballistic regime ($\gamma\ll 1$) or making use of external flows and geometries to direct swimmers.

\subsubsection{Ballistic limit ($\gamma\to0$)}\label{sec:actball}
The asymptotic expansions in the ballistic limit are more complicated, and rely on careful splitting of the integration range to tame divergences. With all details shown in the Appendix, we obtain the following leading-order results:
\begin{align}
	\langle r^2 \rangle_{a}&=U^2T_d^2\left(1-\frac{16}{3}\gamma+\mathcal{O}(\gamma^{3/2})\right),\nonumber\\
	&\text{(non-reacting)}\label{eq:ball1}\\
	\langle r^2 \rangle_{a}&=U^2T_d^2\left(1-2\sqrt{\pi}\sqrt{\gamma}+\mathcal{O}(\gamma\log\gamma)\right),\nonumber\\
	&\text{(slow reaction)}\label{eq:ball2}\\
	\langle r^2 \rangle_{a}&=U^2T_d^2\left(1-4\sqrt{2\pi}\sqrt\gamma+\mathcal{O}(\gamma\log\gamma)\right).\nonumber\\
	&\text{(fast reaction)}\label{eq:ball3}
\end{align}

Once again, we observe the same hierarchy among  the three models, with the non-reacting swimmer exhibiting the smallest decrease in range compared to a classical colloid, in contrast with a fast reacting swimmer with the same lifetime $T_d$. Note that in this limit not only the coefficient but also the leading-order scaling varies between the models.

We obtain therefore that in both the ballistic and diffusive limit there exists a hierarchy among the three models. The mean squared displacement for a given value of $\gamma$ is always largest for the non-reacting swimmer, followed by the slowly reacting and finally the fast reacting colloid. This may be explained by considering the decay behaviour in 
Fig.~\ref{fig:rcomp}. Since the decay rate of the non-reacting swimmer is accelerating, it is only significantly smaller than its original size for a comparatively short proportion of its total lifetime. Since rotational diffusion is strongest for particles of small radius, this means that it is comparatively weakly affected by the enhancement in rotational diffusion. In contrast, colloids decaying according the other two models experience strong rotational diffusion for a significantly longer proportion of their lifetime, leading to less directed motion and smaller overall displacement. In Fig.~\ref{fig:hists} and \ref{fig:scatter} we illustrate this further using results from our numerical simulations.

\subsection{Computational results}\label{sec:Numerics}

\subsubsection{Validation of the method}

\begin{figure}[t]
	\centering
	\includegraphics[width=0.5\columnwidth]{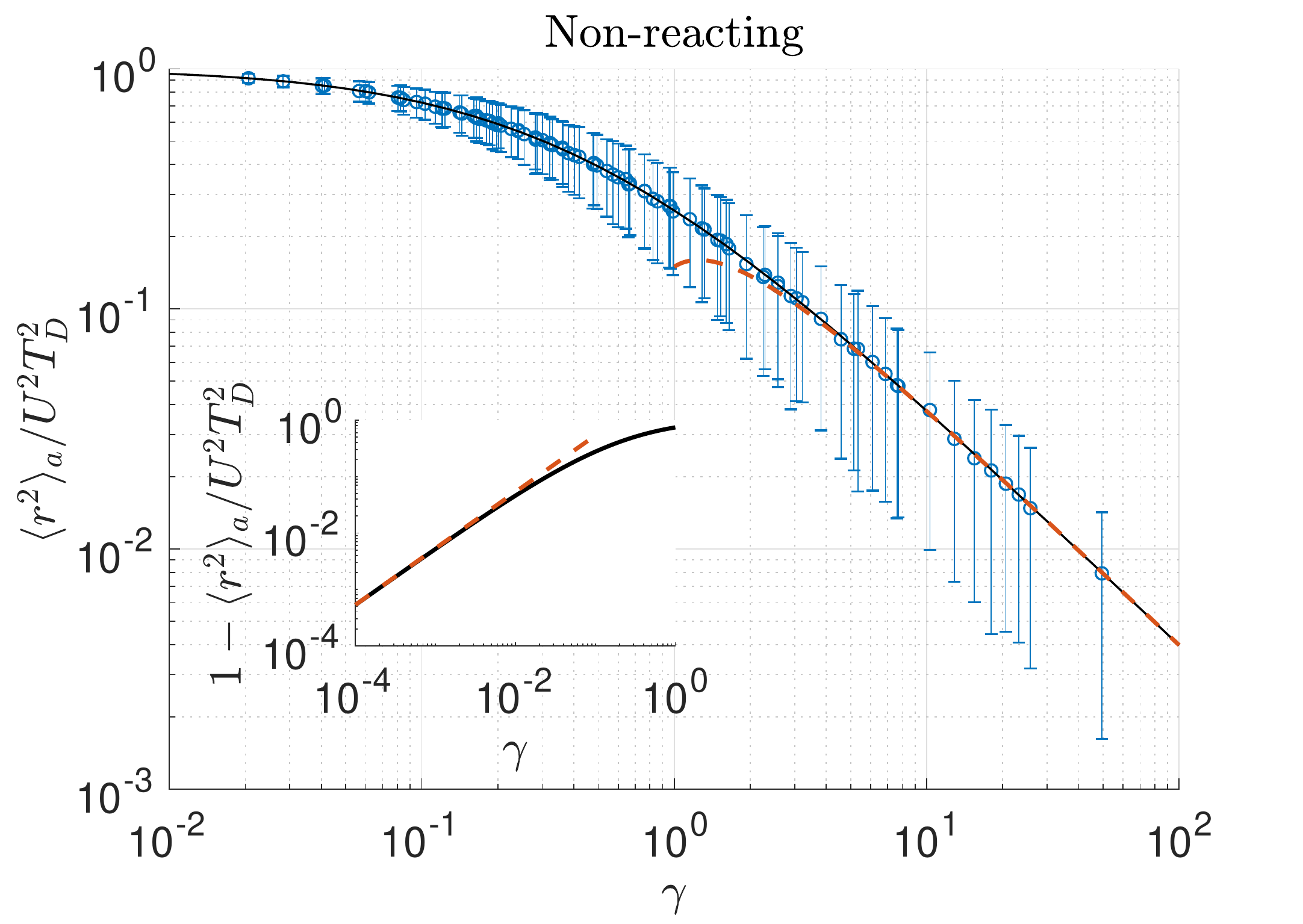}
	\caption{Normalised active mean squared displacement as a function $\gamma$ for the non-reacting model. The solid black line corresponds to direct numerical integration of Eq.~\eqref{eq:amsd1}, while the dashed orange line is our theoretical prediction in Eq.~\eqref{eq:actmsd1} for the large $\gamma$ limit. Each scatter point represents the mean and one standard deviation obtained from $10^3$ Monte-Carlo simulations of the associated Langevin equations. Inset:  the small $\gamma$ behaviour, comparing Eq.~\eqref{eq:amsd1} (solid black) with the asymptotic solution Eq.~\eqref{eq:ball1} (dashed orange).}\label{fig:amsd1}
\end{figure}

\begin{figure}[t]
	\centering
	\includegraphics[width=0.5\columnwidth]{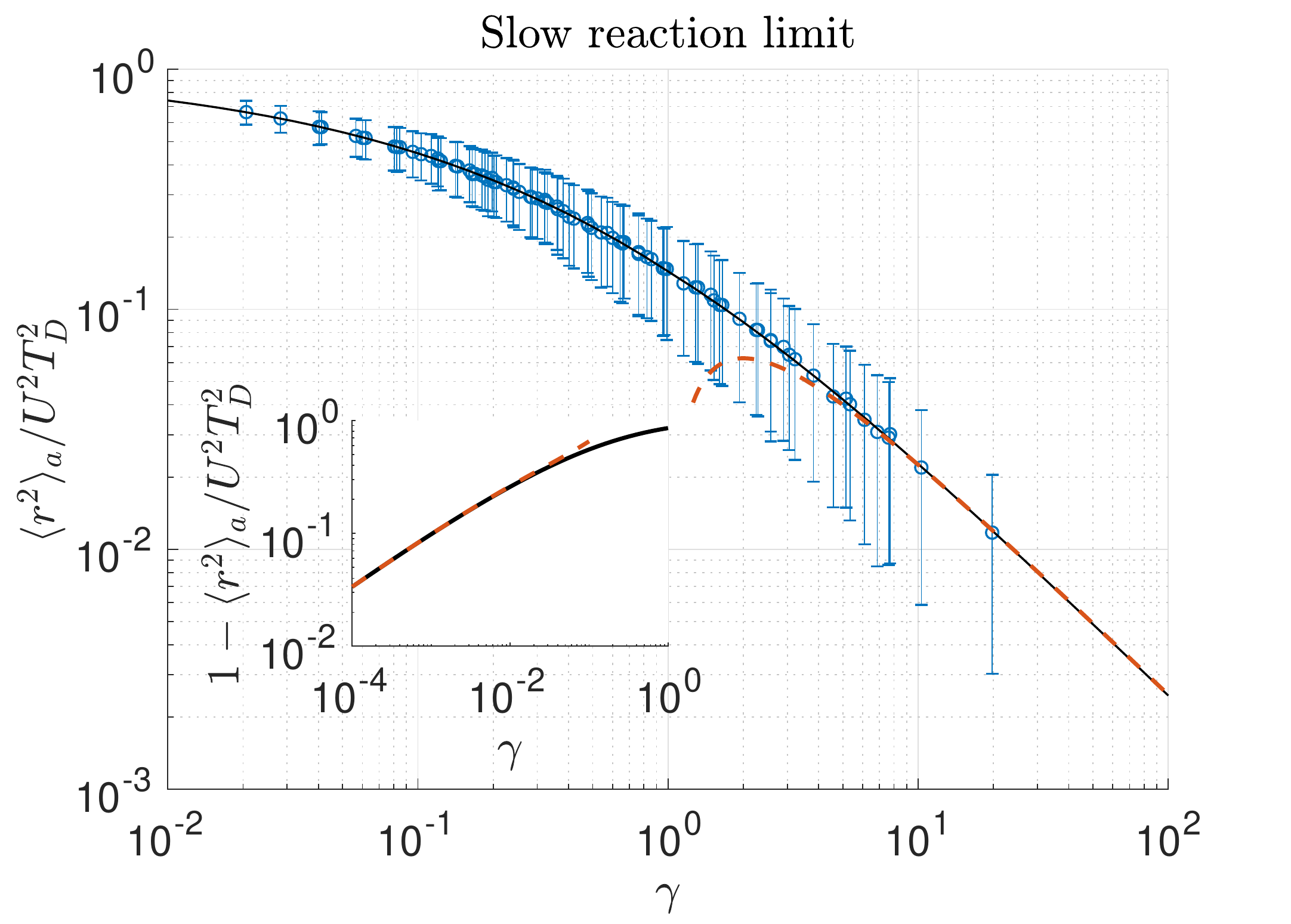}
	\caption{Normalised active mean squared displacement against $\gamma$ for the slow reaction limit of the reacting model. The solid black line corresponds to direct numerical integration of Eq.~\eqref{eq:amsd2}, the dashed orange lines to the theoretical predictions of Eq.~\eqref{eq:actmsd2} and Eq.~\eqref{eq:ball2}, and the scatter points to Monte-Carlo simulations in analogy with Fig. \ref{fig:amsd1}.}\label{fig:amsd2}
\end{figure}

\begin{figure}[t]
	\centering
	\includegraphics[width=0.5\columnwidth]{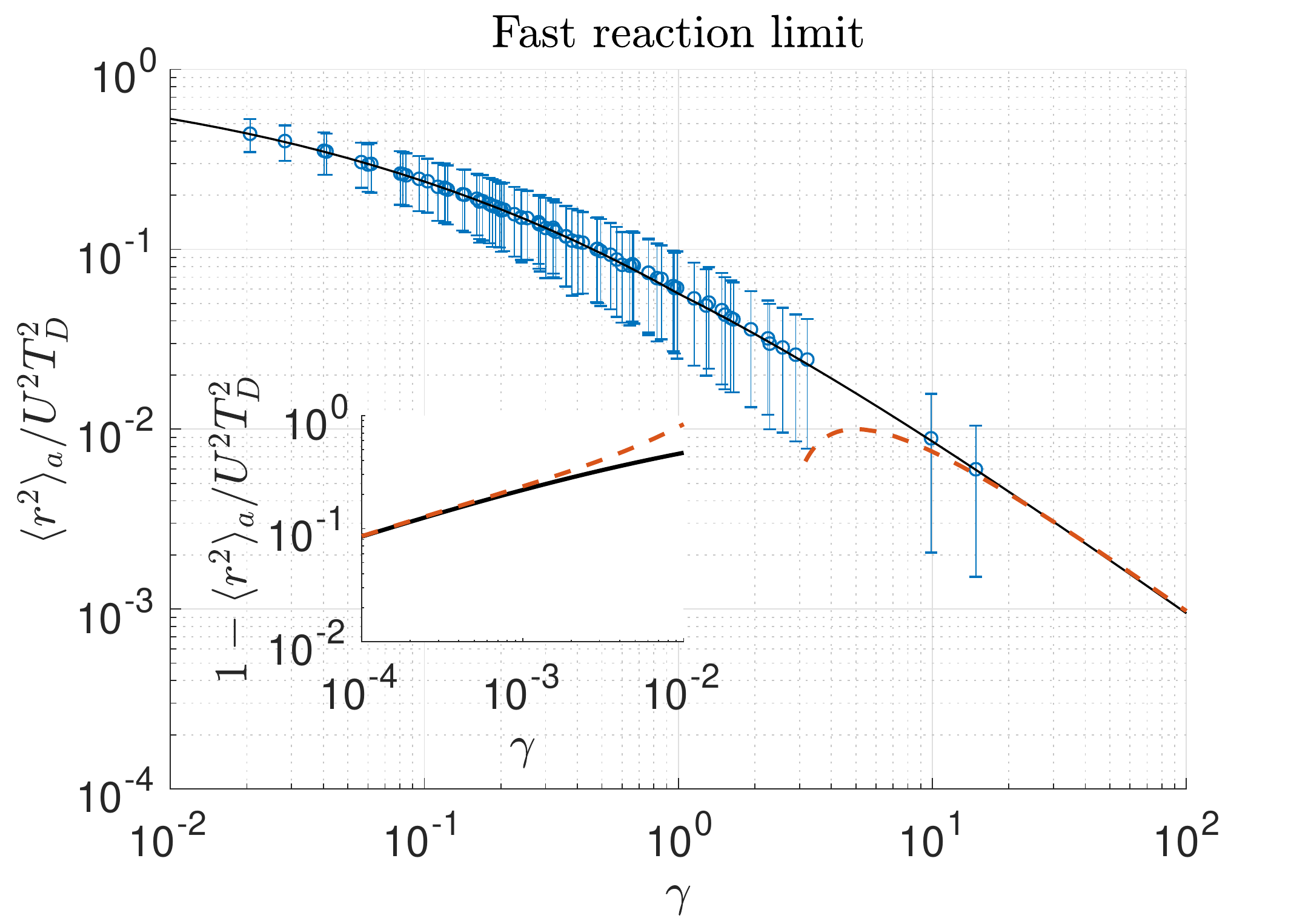}
	\caption{Normalised active mean squared displacement against $\gamma$ for the slow reaction limit of the reacting model. The solid black line corresponds to direct numerical integration of Eq.~\eqref{eq:amsd3}, the dashed orange lines to the theoretical predictions of Eq.~\eqref{eq:actmsd3} and Eq.~\eqref{eq:ball3}, and the scatter points to Monte-Carlo simulations in analogy with Fig. \ref{fig:amsd1}.}\label{fig:amsd3}
\end{figure}

In order to test our theoretical approach, we perform direct numerical integrations of our integral expressions for the active mean squared displacement in Eqs.~\eqref{eq:amsd1}-\eqref{eq:amsd3}. We compare them with Monte-Carlo simulations of the associated Langevin dynamics to assert its validity, and subsequently  with our analytical predictions for the asymptotic behaviour. The results are shown in Figs.~\ref{fig:amsd1}, \ref{fig:amsd2} and \ref{fig:amsd3} for the non-reacting, slowly reacting and fast reacting models respectively. Since the large $\gamma$ limit corresponds to strong rotational diffusion and long lifetimes, the Monte Carlo simulations necessitate very small time steps and very long run times. Depending on the model, such simulations therefore become prohibitively expensive even for moderate values of $\gamma$. Since rotational diffusion is strongest for small colloids, this effect is most pronounced for the fast reacting swimmer whose rate of dissolution is decreasing since this swimmer spends the longest proportion of its lifetime in this regime. Conversely, the non-reacting swimmer is the least expensive to simulate.

As can be seen in Fig.~\ref{fig:amsd1}, we obtain  excellent agreement between the Langevin dynamics and the predicted mean-squared displacement for a wide range of $\gamma$ values. In the diffusive limit ($\gamma\gg 1$), the next-to leading order asymptotics agree extremely well with the exact result down to $\gamma=\mathcal{O}(1)$ on a log-log scale. In the ballistic limit, divergences begin to appear at  $\gamma=\mathcal{O}(10^{-1})$. Similar conclusions hold for the slowly reacting swimmer, as shown in Fig.~\ref{fig:amsd2}. In the case of the fast reacting swimmer, shown in Fig.~\ref{fig:amsd3}, the active mean squared displacement is a less smooth function of $\gamma$, leading to stronger diversion from the asymptotic expressions.

\subsubsection{Distribution of spread}

\begin{figure}[h!]
	\centering
	\begin{subfigure}{\columnwidth}
			\includegraphics[width=.5\columnwidth]{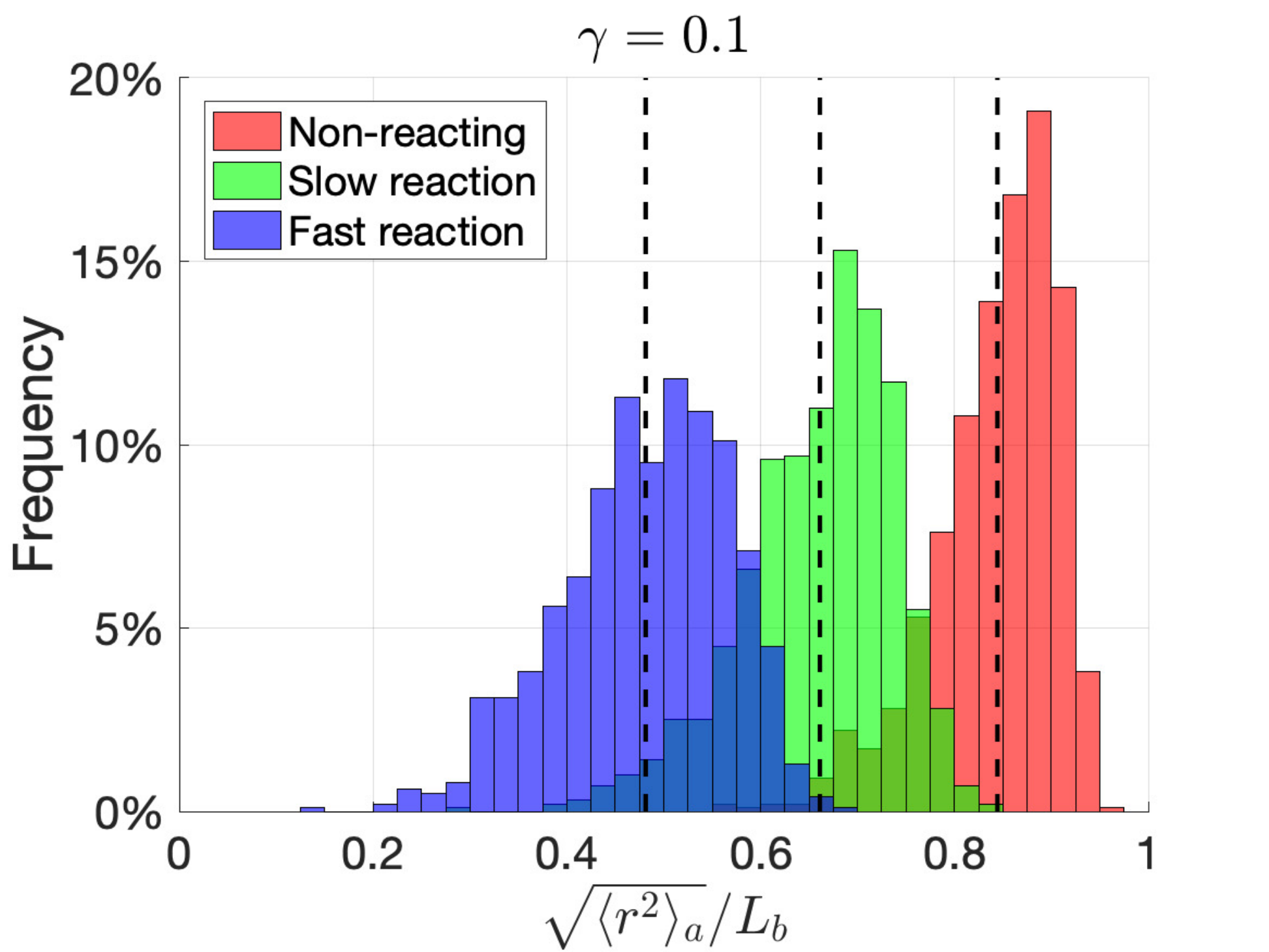}\label{fig:hists1}
	\end{subfigure}\\
	\begin{subfigure}{\columnwidth}
			\includegraphics[width=.5\columnwidth]{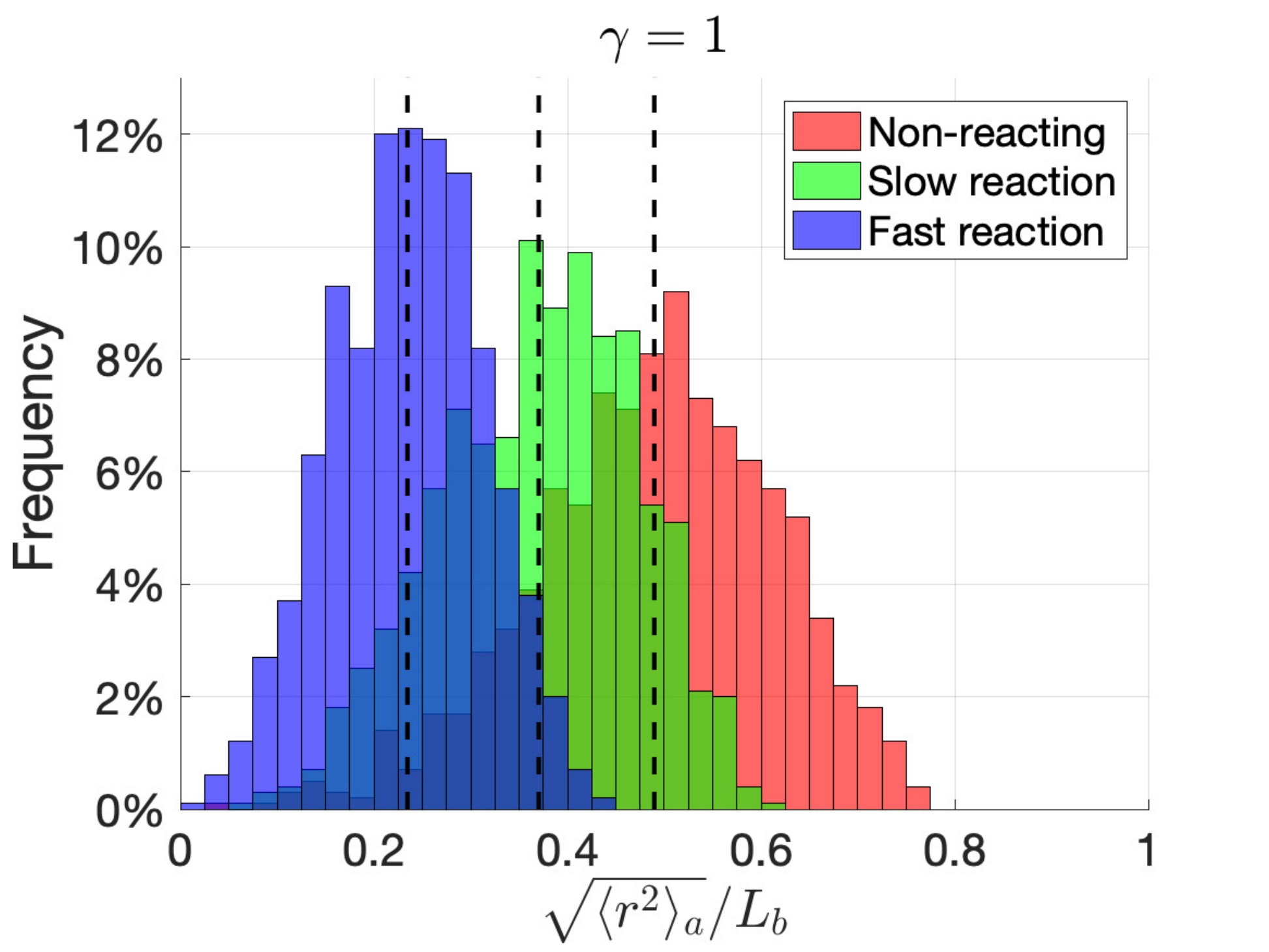}\label{fig:hists2}	
	\end{subfigure}\\
	\begin{subfigure}{\columnwidth}
			\includegraphics[width=.5\columnwidth]{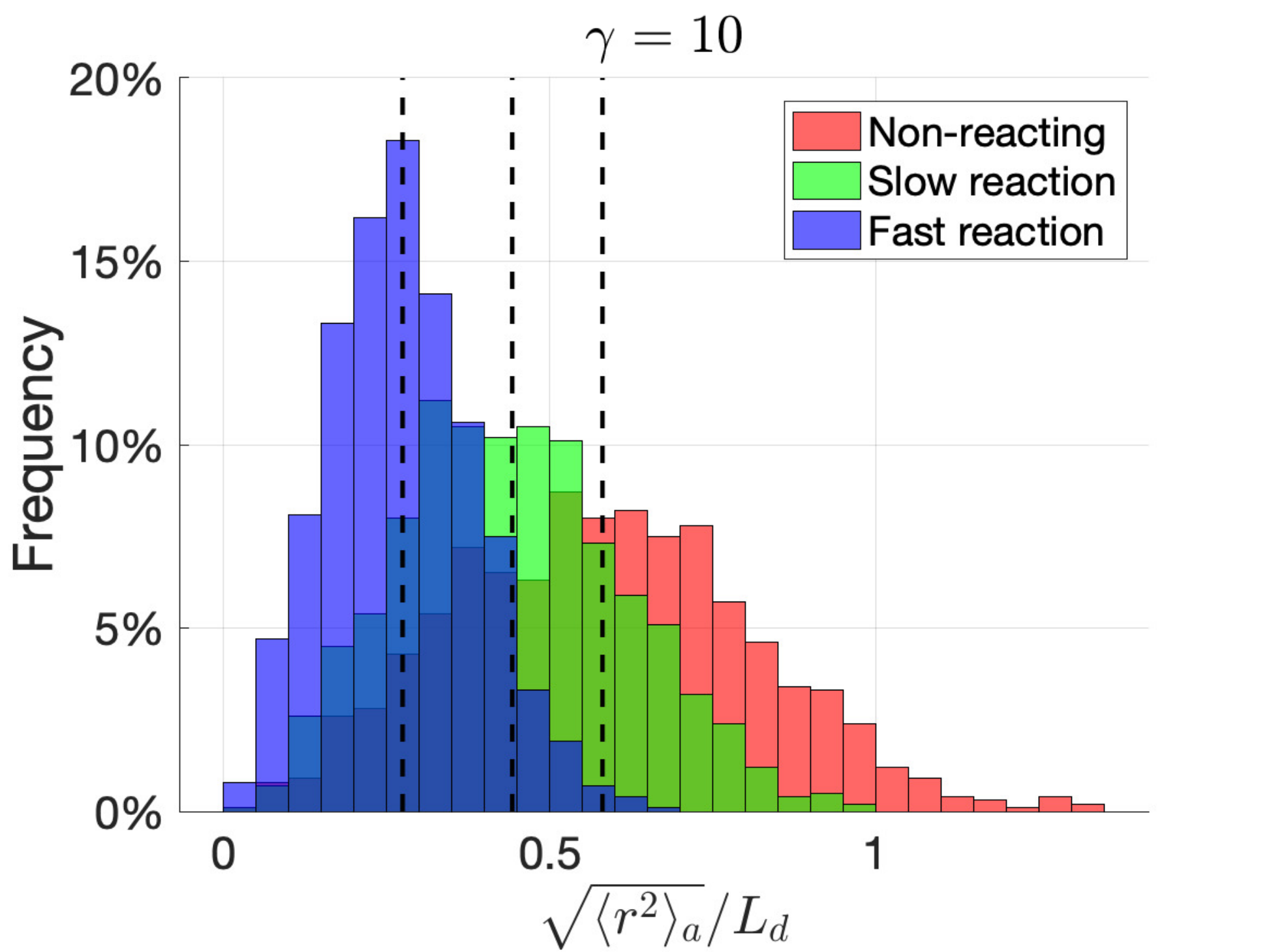}\label{fig:hists3}	
	\end{subfigure}
	\caption{Histograms illustrating the distribution of root mean squared displacement from the initial position for different values of $\gamma$, scaled by the ballistic length scale $L_b=UT_d$ for $\gamma=0.1$ and $\gamma=1$, and the diffusive length scale $L_d=L_b/\sqrt{\gamma}$ for $\gamma=10$. Each histogram is generated from $10^3$ Monte Carlo simulations. Dashed lines indicate sample means.}\label{fig:hists}
\end{figure}

\begin{figure}[h]
	\centering
	\includegraphics[width=.5\columnwidth]{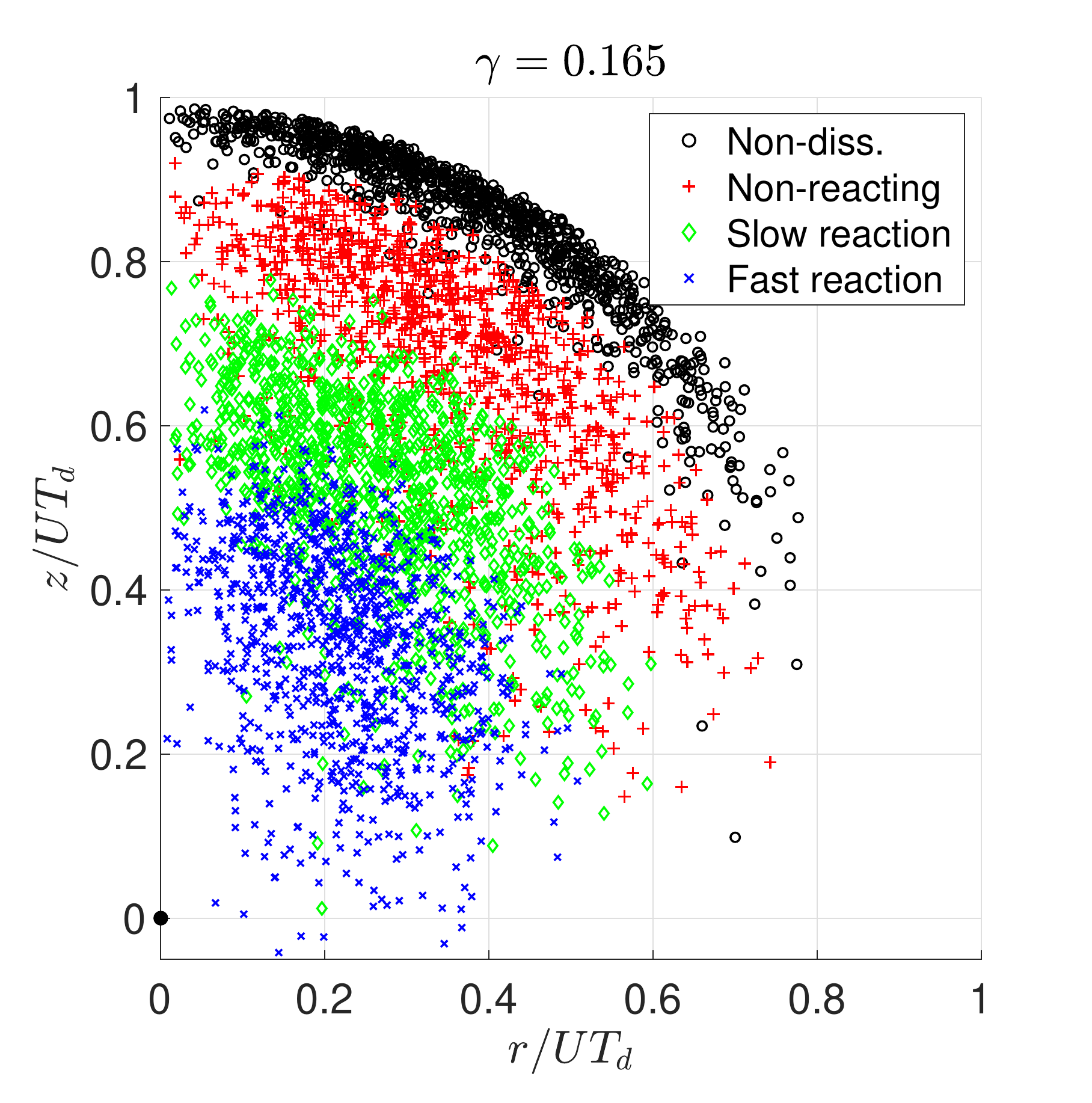}
	\caption{Cloud scatter plot of lateral displacement, $r_\perp=\sqrt{x^2+y^2}$, 
	vs.~vertical displacement, $z$, of $10^3$ Monte Carlo simulations in the weakly ballistic regime for our three models compared to the non-dissolving case. All simulations are started at the coordinate origin (filled circle) with initial orientation vertically upwards in Cartesian coordinates $(x,y,z)$. Symbols indicate positions of the colloids at time of disappearance. The non-dissolving data points are generated by initialising a simulation with a given rotational diffusivity $D_r$ and terminating after a time $T$ such that $TD_r=\gamma$. Lengths are scaled by the ballistic length scale $UT_d$.}\label{fig:scatter}
\end{figure}

From these Monte-Carlo simulations, we can deduce further information regarding the spread of particle trajectories. As predicted in \S\ref{sec:act_matmod}, a hierarchy between the models is revealed that applies for a wide range of values of $\gamma$, covering both the ballistic and the diffusive regime. This is illustrated in Fig.~\ref{fig:hists}, where we show histograms of root-mean-square displacement distributions. For equal values of $\gamma$, the non-reacting model consistently produces the largest displacement. The distribution is strongly peaked for small $\gamma$ (ballistic), but spreads as $\gamma$ shifts to larger values. This may be attributed to the general shift towards diffusion. Contrastingly however, the distribution of the fast-reacting colloids is spread rather widely even in the ballistic regime and in fact peaked much more strongly in the diffusive regime than both the non-reacting and the slowly reacting particles, whose distribution lies between the two others. This is indicative of fast-reacting dissolution fostering diffusive behaviour independent of the parameter $\gamma$.  

In order to further illustrate this point, we examine the lateral spread of colloid trajectories in the weakly ballistic regime. In Fig.~\ref{fig:scatter}, we plot the final positions of colloids with identical initial orientations, including   non-dissolving particles for comparison. A clear stratification between the models is visible with non-dissolving colloids being closely confined to a spherical cap on the one extreme, and fast reacting colloids in a near-spherical diffusive cloud close to the origin. These also exhibit the smallest absolute lateral spread, while the classical colloids are the most spread out. However, the average angular spread is similar between the models.

\section{Discussion}\label{sec:Discussion}

In this paper we provide two fundamental models for the dissolution and stochastic dynamics of self-propelled artificial microswimmers. Inspired by recent experimental realisations, we seek to identify the swimmer decay rates and their influence on translational and rotational diffusivity, and in turn analyse both theoretically and numerically how changes in these modify the distribution of swimmer trajectories. We identify a new dimensionless parameter, $\gamma$ defined as the product of lifetime and initial rotational diffusivity, that classifies colloids with finite lifetime into `ballistic' and `diffusive' types independent of the dissolution process, and study the differences between our dissolution models in three distinct limits for various values of this parameter. We find that for a given value of $\gamma$, particles dissolving in the absence of a reaction behave the most ballistic, whereas colloids reacting at high Damk\"ohler number, defined as the ratio of fuel reactivity and diffusive replenishment, behave the most diffusively. We find that this is due to increasing and decreasing dissolution rates respectively for the different models. Furthermore we derive asymptotic expressions of their mean squared displacement for both small and large values of $\gamma$, and perform extensive Monte Carlo simulations to validate our theoretical results and derive more information about the distribution of spread.

Under experimental conditions, Damk\"ohler numbers of more than about 10 are often very difficult to realise. However, this does not really constrain the applicability of our fast-reacting model, since we only require $t_f/\text{Da}^2\ll T_d$ for the expansion to be valid on the scale of dissolution dynamics. Since typically $t_f\ll T_d$ anyway, we find that even Damk\"ohler numbers of order unity are sufficient for this limit. On the other hand, this argument  implies that very small Damk\"ohler numbers are required in the 
slow-reaction asymptotic limit, a situation which might not be realisable experimentally. Note however that we include also the general expression of the decay for arbitrary Damk\"ohler number in Eq.~\eqref{eq:model3diss}, for which computations similar to the ones provided in \S\ref{sec:Numerics} may be performed.

Despite this, not all our models can apply to all kinds of microswimmer designs. Specifically, the non-reacting model might be at odds with phoretic self-propulsion. Therefore this model only describes colloids that propel through different mechanisms, such as magnetic swimmers. Furthermore, our statistical results only hold true for microswimmers that are fully degradable. A Janus colloid with, e.g.,  degradable and inert halves   is not going to exhibit divergent diffusivity since the relevant  length scale  is bounded. Instead such a swimmer would show a decrease in velocity, which if known can be dealt with  in a manner similar to our theoretical approach. In this case, however, the changing geometry of the swimmer    would likely have to be solved for numerically.

Another important problem that remains to be investigated is the influence of directed motion, such as chemotaxis. Breaking the isotropy of orientational dynamics prevents an analytical investigation similar to the one carried out  in this paper since it relies on the result that the directional correlation of a particle decays exponentially. However, we can still address the issue directly in at least  one special case. It was shown recently in Ref.~\cite{tuatulea2018artificial} that artificial colloids perform chemotaxis by adjusting their trajectory by means of rotation, translation in the direction to a chemical gradient, and translation at an angle, each with a coefficient of strength that can be calculated from the surface activity and mobility of the colloid. In the case of uniform surface activity, the only coefficient that is non-zero is the one giving rise to translation in the direction of a chemical gradient. In particular, the rotational dynamics remain unaffected. In that case, the swimmer trajectories behave therefore  just like we describe in our paper, plus a constant velocity displacing the colloid in the direction of the chemical gradient. Furthermore  numerical work will be required to address the full interplay between chemotaxis behaviour and dissolution dynamics.

Before degradable designs may be employed in real-world applications, it will be furthermore necessary to examine the effects of collective dissolution. Since our models are sensitive to the background distribution of fuel and/or solute, the influence of other nearby colloids on their dissolution will be noticeable. It is conceivable that, in analogy with bubbles~\cite{michelin2018collective}, different decay patterns and complex stochastic behaviour emerges. Similar effects may also be triggered by confinement and also warrant further investigation.
	
\begin{acknowledgements}
	This project has received funding from the European Research Council (ERC) under the European Union's Horizon 2020 research and innovation programme  (grant agreement 682754 to EL).
\end{acknowledgements}

\section*{Author contributions}
EL conceived the study, AC developed models and performed computations, all authors contributed to the interpretation and writing of the manuscript.

\section*{Conflicts of interest}
There are no conflicts to declare.

\appendix

\section{Details of the asymptotics for active MSD}
\subsection{Diffusive limit ($\gamma\to\infty$)}\label{app:3}
The general expression for the active mean squared displacement is
\begin{equation}\label{eq:actmsda}
\langle r^2 \rangle_a = 2U^2\int_{0}^{T_d} dt' \int_{0}^{t'}dt'' \exp\left\{{-2\left[\tau_r(t')-\tau_r(t'')\right]}\right\}.
\end{equation}
In the case of the non-reacting swimmer we have $R\approx R_0\sqrt{1-t/T_d}$, and thus
\begin{equation}
\tau_r = D_{r,0}T_d\int_0^{t/T_d} \frac{dt'}{(1- t')^{3/2}}=2\gamma \left(\frac{1}{\sqrt{1-t/T_d}}-1\right).
\end{equation}
We can use this to change integration variables in Eq.~\eqref{eq:actmsda} by setting $x=\tau_r/\gamma$ and obtain
\begin{equation}
\langle r^2 \rangle_a=2U^2T_d^2\int_{0}^{\infty}dx'\int_{0}^{x'}dx''\frac{e^{-2\gamma(x'-x'')}}{(1+x'/2)^3(1+x''/2)^3}.
\end{equation}
This transformation can be interpreted as mathematically equivalent to the motion of a non-dissolving colloid with constant rotational diffusivity and algebraically decaying velocity. We switch variables again to
\begin{equation}
\begin{aligned}
y'&=x',\\
y''&=x'-x''.
\end{aligned}
\end{equation}
and obtain
\begin{widetext}
\begin{equation}
\langle r^2\rangle_a =2 U^2T_d^2 \int_{0}^{\infty}dy'\int_{0}^{y'}dy'' e^{-2\gamma y''}\left(1+\frac{y'}{2}\right)^{-3}\left(1+\frac{y'-y''}{2}\right)^{-3}.
\end{equation}
\end{widetext}
It is then possible to write the $y''$-integral in terms of auxiliary Gamma functions. These may be expanded in the limit $\gamma\to\infty$ to give
\begin{widetext}
\begin{equation}
\langle r^2\rangle_a =U^2T_d^2 \int_{0}^{\infty}dy'   \frac{64}{\gamma (2 + y')^6}+ \frac{96}{\gamma^2 (2 + y')^7}- \frac{8 e^{-2 \gamma y'}}{\gamma (2 + y')^3}+\mathcal{O}(\gamma^{-3}).
\end{equation}
\end{widetext}
The first two terms can be evaluated directly, while the last one may be expanded using Watson's lemma. We find that
\begin{equation}
\langle r^2\rangle_a =U^2T_d^2\left(\frac{2}{5\gamma}-\frac{1}{4\gamma^2}+\mathcal{O}\left(\gamma^{-3}\right)\right),
\end{equation}
which is the same as Eq.~\eqref{eq:actmsd1}.

The case of a slowly reacting swimmer can be solved in a very similar fashion. This time we have
\begin{equation}
x=\frac{1}{2}\left(\frac{1}{(1-t/T_d)^2}-1\right).
\end{equation}
It follows that the active part of the mean squared displacement may be written as
\begin{equation}
\langle r^2 \rangle_{a}= 	2U^2T_d^2\int_{0}^{\infty}dx'\int_{0}^{x'}dx''\frac{e^{-2\gamma(x'-x'')}}{(1+2x')^{3/2}(1+2x'')^{3/2}} .
\end{equation}
Developing an asymptotic expansion as before we get
\begin{widetext}
\begin{align}
\langle r^2\rangle_a &=U^2T_d^2 \int_{0}^{\infty}dy'   \frac{1}{\gamma (1 + 2y')^3}+ \frac{3}{2\gamma^2 (1 + 2y')^4}- \frac{ e^{-2 \gamma y'}}{\gamma (1 + 2y')^{3/2}}+\mathcal{O}(\gamma^{-3})\nonumber\\
&=U^2T_d^2\left(\frac{1}{4\gamma}-\frac{1}{4\gamma^2}+\mathcal{O}\left(\gamma^{-3}\right)\right),
\end{align}
\end{widetext}
which is Eq.~\eqref{eq:actmsd2}.

Finally, for the fast reacting swimmer we have
\begin{equation}
x=\frac{t/T_d}{(1-\sqrt{t/T_d})^2},
\end{equation}
from which we can derive that
\begin{equation}
\langle r^2 \rangle_{a}=2U^2T_d^2\int_{0}^{\infty}dx'\int_{0}^{x'}dx''\frac{e^{-2\gamma(x'-x'')}}{(1+\sqrt{x'})^{3}(1+\sqrt{x''})^{3}}.
\end{equation}
In this case it is easier to interchange the integrals as $\int_{0}^{\infty}dx'\int_{0}^{x'}dx''=\int_{0}^{\infty}dy''\int_{y''}^{\infty}dy'$ and perform the $y'$-integral first. The resulting expression produced by Wolfram Mathematica 11 contains 1692 terms, but may again be expanded and simplified significantly upon the application of Watson's lemma, giving
\begin{align}
\langle r^2 \rangle_{a}&=U^2T_d^2\int_{0}^{\infty}dy'' e^{-2\gamma y''}\left(\frac{1}{5}-y''+\mathcal{O}\left(y''^{3/2}\right)\right)\\
&=U^2T_d^2\left(\frac{1}{10\gamma}-\frac{1}{4\gamma^2}+\mathcal{O}\left(\gamma^{-5/2}\right)\right),
\end{align}
as claimed in Eq.~\eqref{eq:actmsd3}.

\subsection{Ballistic limit ($\gamma\to 0$)}\label{app:4}
First, the non-reacting swimmer. We have
\begin{equation}
\langle r^2 \rangle _a = 2U^2T_d^2\int_{0}^{\infty}dx\int_{0}^{x}dy \frac{e^{-2\gamma(x-y)}}{(1+x/2)^3(1+y/2)^3},
\end{equation}
and are interested in the limit $\gamma\to 0$. We set $U^2T_d^2=1$ to keep the notation clean. Since the denominator decays rapidly enough at $\infty$ we can Taylor expand the exponential to pick up the two leading-order contributions to the integral.
\begin{align}
\langle r^2 \rangle _a &= \int_{0}^{\infty}dx\int_{0}^{x}dy \frac{2-4\gamma(x-y)+\dots}{(1+x/2)^3(1+y/2)^3}\\
&=1-\frac{16}{3}\gamma+o(\gamma),
\end{align}
which is Eq.~\eqref{eq:ball1}.

For the slowly reacting swimmer we have
\begin{equation}
\langle r^2 \rangle _a = 2\int_{0}^{\infty}dx\int_{0}^{x}dy \frac{e^{-2\gamma(x-y)}}{(1+2x)^{3/2}(1+2y)^{3/2}}.
\end{equation}
Because of the slower decay, it is necessary to divide and conquer from the start. We set $z=x-y$ and note that $\int_{0}^{\infty}dx\int_{0}^{x}dy=\int_{0}^{\infty}dz\int_{z}^{\infty}dx$. Upon performing the inner integral we have
\begin{equation}
\langle r^2 \rangle _a = \int_{0}^{\infty}dz \frac{e^{-2\gamma z}}{1 + \sqrt{1 + 2 z} + z (2 + \sqrt{1 + 2 z})}.
\end{equation}
We define $\delta$ such that $1\ll\delta\ll\gamma^{-1}$ and split the integral into
\begin{align}
I_1=\int_{0}^{\delta}dz \frac{e^{-2\gamma z}}{1 + \sqrt{1 + 2 z} + z (2 + \sqrt{1 + 2 z})},\nonumber\\
I_2=\int_{\delta}^{\infty}dz \frac{e^{-2\gamma z}}{1 + \sqrt{1 + 2 z} + z (2 + \sqrt{1 + 2 z})}.
\end{align}
Upon expanding the exponential in $I_1$ and taking $\delta\to\infty$ we have
\begin{equation}
I_1=1+(2-2\log 2)\gamma+\mathcal{O}(\gamma^2)+\text{terms depending on }\delta.
\end{equation}
Meanwhile, we rescale $z\to\gamma z$ in $I_2$ and expand the denominator for small $\gamma$.
\begin{equation}
I_2=\int_{\gamma\delta}^{\infty}dze^{-2z}\left(\frac{\gamma^{1/2}}{\sqrt{2}z^{3/2}}-\frac{\gamma}{z^2}+\dots\right).
\end{equation}
Performing the integral and taking the limit $\delta\to 0$ we arrive at
\begin{align}
I_2=-2\sqrt{\pi}\gamma^{1/2}-2\gamma\log\gamma+\left(2-2\gamma_e-2\log 2\right)\gamma\nonumber\\
+o(\gamma)+\text{terms depending on }\delta,
\end{align}
where $\gamma_e$ is the Euler-Mascheroni constant. Since $\delta$ is arbitrary, the divergent terms in both integrals must cancel. In summary, we have for the slowly reacting swimmer that
\begin{equation}
\langle r^2 \rangle _a=1-2\sqrt{\pi}\gamma^{1/2}-2\gamma\log\gamma+\left(4-2\gamma_e-4\log 2\right)\gamma+o(\gamma),
\end{equation}
which is Eq.~\eqref{eq:ball2}.

Finally, for the fast reacting swimmer we have
\begin{equation}
\langle r^2 \rangle _a = 2\int_{0}^{\infty}dx\int_{0}^{x}dy \frac{e^{-2\gamma(x-y)}}{(1+\sqrt{x})^3(1+\sqrt{y})^3}.
\end{equation}
This time there is no closed-form expression for the inner integral, forcing us to split both integrals in two domains. We define $\delta$ as before and write
\begin{equation}
\langle r^2 \rangle _a = \underbrace{\int_{0}^{\delta}dx\int_{0}^{x}dy}_{I_1} + \underbrace{\int_{\delta}^{\infty}dx\int_{0}^{x}dy}_{I_2} \frac{2e^{-2\gamma(x-y)}}{(1+\sqrt{x})^3(1+\sqrt{y})^3}.
\end{equation}
The first part, $I_1$, is straightforward to do once the exponential is expanded and yields
\begin{equation}
I_1=1-\frac{296}{3}\gamma+\mathcal{O}(\gamma^2)+\text{terms depending on }\delta.
\end{equation}
To perform $I_2$ we write
\begin{equation}
I_2=\int_{\delta}^{\infty}dx \frac{2e^{-2 \gamma x}}{(1+\sqrt{x})^3}\underbrace{\int_{0}^{x}dy\frac{e^{2\gamma  y}}{(1+\sqrt{y})^3}}_{J(x)},
\end{equation}
and split the range of $J(x)$ again with the goal to obtain an expansion valid for small $\gamma$. Defining $\delta_1$, $J_1$ and $J_2$ in a similar fashion, we find
\begin{equation}
J_1=1+10\gamma+\mathcal{O}(\gamma^2)+\text{terms depending on }\delta_1,
\end{equation}
whereas for $J_2$ we have
\begin{align}
J_2=&\frac{3e^{2\gamma x}}{x}-\frac{2e^{2\gamma x}}{\sqrt{x}}+2\sqrt{2\pi}\gamma^{1/2}\text{Erfi}\left(\sqrt{2\gamma x}\right)\nonumber\\
&-6\gamma\text{Ei}\left(2\gamma x\right)+2\gamma\log \gamma+\gamma\left(6\gamma_e-6+6\log 2\right)\nonumber\\
&+o(\gamma)+\text{terms depending on }\delta_1,
\end{align}
where $\text{Erfi}(z)=\text{Erf}(iz)/i$ and $\text{Ei}(z)=-\int_{-z}^{\infty}e^{-t}/t\, dt$. Combining these allows us to write
\begin{align}
I_2=\int_{\gamma\delta}^{\infty}dz \frac{2\gamma^{1/2}e^{-2z}}{z^{3/2}}-\frac{\gamma}{z^2}\left(6e^{-2z}+4-4\sqrt{2\pi z}\text{Erfi}\left(\sqrt{2z}\right)\right)\nonumber\\
+o(\gamma).
\end{align}
Expanding as before and combining with $I_1$ we ultimately find that
\begin{align}
\langle r^2 \rangle_{a}=1-4\sqrt{2\pi}\gamma^{1/2}-28\gamma\log\gamma -\gamma\left(\frac{164}{3}+28\gamma_e+60\log 2\right)\nonumber\\
+o(\gamma),
\end{align}
corresponding to Eq.~\eqref{eq:ball3} in the main text.

\bibliography{bibliography}

\end{document}